\documentclass[sigconf, 9pt]{acmart}

\renewcommand\footnotetextcopyrightpermission[1]{}

\usepackage[title]{appendix}
\usepackage{blindtext, graphicx}
\usepackage{algorithm}
\usepackage{algorithmicx}
\usepackage[noend]{algpseudocode}

\usepackage{amssymb}

\usepackage{amsmath,amsfonts} 
\usepackage{braket}
\usepackage[hang,footnotesize,bf,sf]{subfigure}
\usepackage{caption}
\usepackage{comment}
\usepackage{standalone}
\usepackage{tikz}
\usepackage{xcolor}
\usepackage{multirow}
\usepackage{booktabs}
\input{Qcircuit}
\usepackage{nicefrac}
\usepackage{multicol,lipsum}
\usepackage{mathtools, cuted}

\newcommand{\cnot}{\textrm{CNOT}}
\newcommand{\swap}{\textrm{SWAP}}
\providecommand{\min}{\textsc{min}{ }}

\usepackage{fancyhdr}
\usepackage[utf8]{inputenc}
\usepackage[english]{babel}
\fancypagestyle{firstpage}{
  \fancyhf{}
   \fancyhead[C]{\vspace{5pt}\normalsize{CONFIDENTIAL DRAFT: DO NOT CITE, QUOTE, OR DISTRIBUTE.}}
}

\begin{document}
\title{Mapping of Lattice Surgery-based Quantum Circuits on Surface Code Architectures}

\author{\large L. Lao\textsuperscript{1}, B. van Wee\textsuperscript{1}, I. Ashraf\textsuperscript{1}, J. van Someren\textsuperscript{1}, N. Khammassi\textsuperscript{1}, K. Bertels\textsuperscript{1}, C. G. Almudever\textsuperscript{1}\vspace{0.2cm}\\
\textsuperscript{1} \textit{Delft University of Technology, The Netherlands}
}



\begin{abstract}
Quantum error correction (QEC) and fault-tolerant (FT) mechanisms are essential for reliable quantum computing. However, QEC considerably increases the computation size up to four orders of magnitude. Moreover, FT implementation has specific requirements on qubit layouts, causing both resource and time overhead. Reducing spatial-temporal costs becomes critical since it is beneficial to decrease the failure rate of quantum computation.
To this purpose, scalable qubit plane architectures and efficient mapping passes including placement and routing of qubits as well as scheduling of operations are needed. 
This paper proposes a full mapping process to execute lattice surgery-based quantum circuits on two surface code architectures, namely a checkerboard and a tile-based one. We show that the checkerboard architecture is $2$x qubit-efficient but the tile-based one requires lower communication overhead in terms of both operation overhead (up to $\sim 86\%$) and latency overhead (up to $\sim 79\%$ ).
\end{abstract}

\maketitle
\pagestyle{plain}

\section{Introduction}
\label{sec-intro}
By exploiting superposition and entanglement, quantum computing can outperform classical computing while solving certain problems. 
For example, quantum computers can factor large numbers using Shor's algorithm with an exponential speedup over its best classical counterparts~\cite{shor1994algorithms}.
When adopting the circuit model as a computational model, algorithms can be described by quantum circuits consisting of qubits and gates. 
Such a circuit representation is hardware agnostic and assumes, for instance, that any arbitrary interaction between qubits is possible and both qubits and gates are reliable.

However, real quantum processors have specific constraints that must be complied to when executing a quantum algorithm, a procedure for mapping quantum circuits is therefore required. 
One of the main constraints in current quantum experimental platforms is the limited connectivity between qubits. 
A promising qubit structure that is being pursued for many quantum technologies like superconductors~\cite{barends2014superconducting, versluis2016scalable} and quantum dots~\cite{hill2015surface, li2017crossbar}, is a 2D grid architecture that only allows \textit{nearest-neighbour} (NN) interactions. 
Other 2D qubit structures such as the quantum processors from IBM~\cite{IBM17}, Google~\cite{boixo2016characterizing}, and Rigetti~\cite{sete2016functional} have even more restrictive connectivity constraints. 
This means that non-neighbouring or non-connected qubits need to be moved or routed to be adjacent for interacting -i.e. performing a two-qubit gate, resulting in an overhead in the number of operations as well as the execution time (latency) of the circuit.

Placing frequently interacting qubits close to each other combined with efficient routing techniques -e.g. shortest path- can help to reduce the movement overhead. 
In addition, exploiting available parallelism of operations will reduce the overall execution time of the circuit. 
Note that reducing the number of operations and the total circuit latency will be of benefit to decrease the failure rate of computation~\cite{bishop2017quantum,linke2017experimental}. 
Therefore, efficiently mapping quantum circuits on a specific qubit structure, including placement and routing of qubits and scheduling of operations, is necessary for reliable quantum computation. 
Many works have been done to map physical quantum circuits on different qubit structures. ~\cite{metodi2006scheduling, whitney2007automated, dousti12min,yazdani2013quantum, bahreini2015minlp, lye2015determining, wille2016look,farghadan2017quantum} propose algorithms to map physical circuits on quantum processors with 2D NN structures.
~\cite{IBMQISKIT, Zulehner2017efficient, siraichi2018qubit} and~\cite{venturelli2018compiling} respectively focus on IBM and Rigetti processors which both only support interactions on dedicated neighbours. 

Moreover, quantum hardware is error prone, that is, the qubits loose their states (or decohere) extremely fast and quantum operations are faulty.
For instance, superconducting qubits decohere in tens of microseconds~\cite{riste2015detecting} and quantum operations have error rates $\sim0.1\%$~\cite{kelly2015state} compared to $\sim 10^{-15}$ for CMOS based devices. 
Therefore, \textit{quantum error correction} (QEC) and \textit{fault-tolerant} (FT) mechanisms are needed to protect quantum states from errors and make quantum computing FT. 
This is achieved by encoding a \textit{logical} qubit into multiple error prone \textit{physical} qubits and applying FT (logical) operations on such logical qubits~\cite{nielsen2010quantum}. 
However, QEC significantly increases the computation size up to four orders of magnitude. Furthermore, this FT implementation may lead to more and/or different constraints on the encoded logical circuits, e.g., interaction restrictions between two logical qubits. 
Consequently, the mapping of fault-tolerant quantum circuits may become more difficult because it should consider both physical-level and logical-level constraints. 
In addition, it may require the definition of a virtual layer called qubit plane architecture to provide scalable management of qubits and support fast execution of fault-tolerant operations.

Several papers~\cite{balensiefer2005quale, dousti2013leqa, dousti2014squash, ahsan2015architecture, heckey2015compiler, lin2015paqcs} have discussed how to map FT quantum circuits onto 2D quantum architectures based on concatenated codes such as Steane code. 
However, not many papers focus on surface code (SC)~\cite{bravyi1998quantum}, currently one of the most promising QEC codes. 
~\cite{paler2016synthesis,paler2017fault, paler2017online} optimize quantum circuits based on defect surface codes in terms of geometrical volume defined by the product of \# qubits and \# gates (or time) of the circuit. 
~\cite{javadi2017optimized} evaluates both planar and defect surface codes in terms of qubit resources and circuit latency. 
However, they assume two-qubit gates ($\cnot$) between two planar qubits can be performed transversally, which is an over-optimistic assumption given the limited connectivity in current quantum technologies. 
Fortunately, a technique called lattice surgery~\cite{horsman2012surface,landahl2014quantum} can be used to perform a two-qubit gate between two planar qubits in a 2D NN architecture. 
Nevertheless, the mapping of quantum circuits based on lattice surgery and the required qubit plane architecture have been hardly researched.
\cite{horsman2012surface} introduces a scalable qubit architecture for efficiently supporting lattice surgery-based two-qubit gates. 
\cite{herr2017optimization} proves that the optimization of lattice surgery-based quantum circuits on its geometrical volume is NP-hard.

This paper will focus on the mapping of lattice surgery-based quantum circuits onto surface code qubit architectures. 
The contributions of this paper are the following:

\begin{itemize}
    
    \item We derive the logical-level constraints of the mapping process when the lattice surgery is used to perform FT operations on planar surface codes. 
    We further provide the quantification of these logical operations, which are used for the mapping passes. 
    \item Based on the qubit plane architecture presented in \cite{horsman2012surface}, we propose two different qubit architectures, namely a checkerboard architecture and a tile-based one, that support lattice surgery-based operations. 
    For the tile-based architecture, we present an approach to fault-tolerantly swap tiles by lattice surgery, which is $3$x faster than a standard $\swap$ operation by $3$ consecutive logical $\cnot$ gates. 
    In addition, we also apply similar techniques to perform a FT CNOT gate between tiles where logical data qubits are not located in the required positions.
    \item We propose a full mapping procedure, including placement and routing of qubits and scheduling of operations, to map FT quantum circuits onto the two presented qubit architectures and evaluate these architectures on their communication overhead. 
    
\end{itemize}

The paper is organized as follows. 
Section~\ref{sec_ft} introduces the basics of FT quantum computing. 
We introduce two qubit plane architectures of interest in Section~\ref{sec-arch} followed by the proposed mapping passes in Section~\ref{sec-mapping}. 
The evaluation metrics and benchmarks are shown in Section~\ref{sec-metric}. 
The experimental results are discussed in Section~\ref{sec-result}. Section~\ref{sec-conc} concludes.

\section{FT quantum computing}\label{sec_ft}

Like in classical computing, quantum computing is also built on a two-level system named qubit. 
A qubit however can be in a superposition of states $\Ket{0}$ and $\Ket{1}$:  \(\Ket{\psi}=\alpha\Ket{0}+\beta\Ket{1}\) where $\alpha$ and $\beta$ are complex numbers. 
Quantum states can be transformed by performing quantum operations on them. Commonly-used quantum gates include single-qubit gates, such as Pauli-$X$, -$Y$, -$Z$, Hadamard ($H$), $S$ and $T$, and two-qubit gates, such as the Controlled-NOT ($\cnot$) and $\swap$.
In a $\cnot$ gate, the target qubit `T' is flipped only if the control qubit `C' is $\ket{1}$. 
A $\swap$ gate interchanges the states of two qubits and can be implemented by three consecutive $\cnot$ gates. 
The gate set $\{H, S, T, \cnot\}$ is one of the most popular universal sets of quantum gates, meaning that any arbitrary quantum gate can be approximated within a particular precision by a finite sequence of those gates. 
Any quantum algorithm can be described by a quantum circuit which consist of qubits and quantum gates.

\subsection{Quantum error correction}
\begin{figure}[tb!]
\centerline{
\subfigure[]{\includegraphics[width=1.3in]{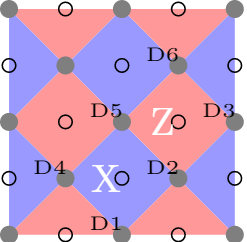}
\label{layout_sc}}\vspace{-2mm}
\subfigure[]{\includegraphics[width=1.5in]{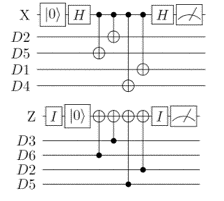}
    \label{esm}}}
\caption{(a) The qubit layout of a surface code, where data qubits are on the vertices (solid circles) and $X$- and $Z$-ancilla qubits are on the purple and pink plaquettes (open circles), respectively. 
(b) Error syndrome measurement circuits for $X$- and $Z$-stabilizers ($X_{D5,D2,D4,D1}$ and $Z_{D6,D3,D5,D2}$). }
\label{sc}
\end{figure}

As mentioned before, quantum systems are error prone so that QEC is required for reliable computation.
The idea of QEC is to encode a \textit{logical} qubit into many \textit{physical} qubits and constantly check the system to detect possible errors.
The number of errors that can be corrected is determined by the code distance $d$ which is defined as the minimum number of physical operations required to perform a logical operation.
Surface code is one of the most promising QEC codes because of its high tolerance to errors (around $1\%$) and its simple 2D structure with only NN interactions as shown in Figure~\ref{layout_sc}. 
It consists of two types of qubits, data qubits (solid circles) for storing computational information, and $X$- or $Z$-ancilla qubits (open circles) used to perform stabilizer measurement. 
The stabilizer measurements are also called error syndrome measurements (ESM) of which circuit description is shown in Figure~\ref{esm}. Note that the $\cnot$ gates are only performed between ancilla qubits and their nearest-neighbouring data qubits. 
We define a SC cycle as the interval between the starting points of two consecutive ESM. 

In surface code, there are two main ways of encoding a single logical qubit, using a \textit{planar} ~\cite{dennis2002topological} or a \textit{defect} approach~\cite{raussendorf2006fault}.
In planar SC, a single lattice is used to encode one logical qubit. In defect SC, a logical qubit is realized by creating defects in a lattice. 
For both codes, an implementable universal set of FT logical operations are initialization (Init) and measurement (MSMT) of qubits, Pauli, $H$, $S$, $T$ and $\cnot$ gates.
However, planar SC requires less physical qubits to encode one logical qubit for the same code distance.
In the near-term implementation of quantum computing, qubits are scarce resources and current quantum technologies are pursuing a realization of planar SC quantum hardware \cite{versluis2016scalable}.
This paper therefore focuses on planar surface code. Note that the FT implementation of defect SC \cite{fowler2012surface, raussendorf2006fault,raussendorf2007fault,raussendorf2007topological} differs from planar SC, leading to different implications on the mapping procedure.

\begin{figure}[tb!]
\centerline{\subfigure[]{\begin{tikzpicture}

\colorlet{colorx}{blue!40!}
\colorlet{colorz}{red!40!}

\colorlet{color0.0}{colorx}
\colorlet{color1.0}{colorz}
\colorlet{udcolor0.0}{colorx}
\colorlet{udcolor1.0}{white}
\colorlet{lrcolor0.0}{white}
\colorlet{lrcolor1.0}{colorz}

\def\dx{0}
    \pgfmathsetmacro{\dx}{1}
\def\dy{0}
    \pgfmathsetmacro{\dy}{1}

\def\Dx{0}
    \pgfmathsetmacro{\Dx}{\dx+1}
\def\Dy{0}
    \pgfmathsetmacro{\Dy}{\dy+1}
    
\def\sftonex{0}
    \pgfmathsetmacro{\sftonex}{0}
\def\sftoney{0}
    \pgfmathsetmacro{\sftoney}{3}
    
\def\sfttwox{0}
    \pgfmathsetmacro{\sfttwox}{3}
\def\sfttwoy{0}
    \pgfmathsetmacro{\sfttwoy}{0}
    
\def\sftldx{0}
    \pgfmathsetmacro{\sftldx}{0}
\def\sftldy{0}
    \pgfmathsetmacro{\sftldy}{0}
    
\def\sftrux{0}
    \pgfmathsetmacro{\sftrux}{3}
\def\sftruy{0}
    \pgfmathsetmacro{\sftruy}{3}

    
    
    
    
\foreach \x in {0,..., \dx}
    \foreach \y in {0, ..., \dy}
    {
        \def\bit{0}
            \pgfmathsetmacro{\bit}{mod((\x+\y),2)}
            \fill[color\bit] (\x+\sftonex, \y+\sftoney) rectangle (\x+1+\sftonex, \y+1+\sftoney);
    
    }

\foreach \x in {0,..., \dx}
    {
        \def\bit{0}
            \pgfmathsetmacro{\bit}{mod((\x+\dy+1),2)}
        \path [fill=udcolor\bit] (\x+\sftonex,\dy+1+\sftoney) to [out=90,in=180] (\x+0.5+\sftonex, \dy+1.5+\sftoney) to [out=0,in=90] (\x+1+\sftonex,\dy+1+\sftoney) --(\x+\sftonex,\dy+1+\sftoney);
        \path [fill=udcolor\bit] (\x+1+\sftonex,0+\sftoney) to [out=-90,in=180] (\x+1.5+\sftonex, -0.5+\sftoney) to [out=0,in=-90] (\x+2+\sftonex,0+\sftoney) --(\x+1+\sftonex,0+\sftoney);
    }

\foreach \y in {0, ..., \dy}
    {
        \def\bit{0}
            \pgfmathsetmacro{\bit}{mod((\dx+1+\y),2)}
        \path [fill=lrcolor\bit] (\dx+1+\sftonex,\y+\sftoney) to [out=0,in=-90] (\dx+1+0.5+\sftonex, \y+0.5+\sftoney) to [out=90,in=0] (\dx+1+\sftonex,\y+1+\sftoney) --(\dx+1+\sftonex,\y+\sftoney);
        \path [fill=lrcolor\bit] (0+\sftonex,\y-1+\sftoney) to [out=180,in=-90] (-0.5+\sftonex, \y-1+0.5+\sftoney) to [out=90,in=180] (0+\sftonex,\y-1+1+\sftoney) --(0+\sftonex,\y-1+\sftoney);
    }

\foreach \x in {0, 1, ..., \dx}
    \foreach \y in {0, 1, ..., \dy}
    {
        \def\bit{0}
            \pgfmathsetmacro{\bit}{mod((\x+\y+1),2)}
            \fill[color\bit] (\x, \y) rectangle (\x+1, \y+1);
    
    }

\foreach \x in {0, 1, ..., \dx}
    {
        \def\bit{0}
            \pgfmathsetmacro{\bit}{mod((\x+\dy),2)}
        \path [fill=udcolor\bit] (\x,\dy+1) to [out=90,in=180] (\x+0.5, \dy+1.5) to [out=0,in=90] (\x+1,\dy+1) --(\x,\dy+1);
        \path [fill=udcolor\bit] (\x-1,0) to [out=-90,in=180] (\x+0.5-1, -0.5) to [out=0,in=-90] (\x+1-1,0) --(\x-1,0);
    }

\foreach \y in {0, 1, ..., \dy}
    {
        \def\bit{0}
            \pgfmathsetmacro{\bit}{mod((\dx+\y),2)}
        \path [fill=lrcolor\bit] (\dx+1,\y) to [out=0,in=-90] (\dx+1+0.5, \y+0.5) to [out=90,in=0] (\dx+1,\y+1) --(\dx+1,\y);
        \path [fill=lrcolor\bit] (0,\y+1) to [out=180,in=-90] (-0.5, \y+1.5) to [out=90,in=180] (0,\y+2) --(0,\y+1);
    }

\foreach \x in {0, 1, ..., \dx}
    \foreach \y in {0, 1, ..., \dy}
    {
        \def\bit{0}
            \pgfmathsetmacro{\bit}{mod((\x+\y+1),2)}
            \fill[color\bit] (\x+\sftrux, \y+\sftruy) rectangle (\x+1+\sftrux, \y+1+\sftruy);
    
    }

\foreach \x in {0, 1, ..., \dx}
    {
        \def\bit{0}
            \pgfmathsetmacro{\bit}{mod((\x+\dy),2)}
        \path [fill=udcolor\bit] (\x+\sftrux,\dy+1+\sftruy) to [out=90,in=180] (\x+0.5+\sftrux, \dy+1.5+\sftruy) to [out=0,in=90] (\x+1+\sftrux,\dy+1+\sftruy) --(\x+\sftrux,\dy+1+\sftruy);
        \path [fill=udcolor\bit] (\x-1+\sftrux,0+\sftruy) to [out=-90,in=180] (\x+0.5-1+\sftrux, -0.5+\sftruy) to [out=0,in=-90] (\x+1-1+\sftrux,0+\sftruy) --(\x-1+\sftrux,0+\sftruy);
    }

\foreach \y in {0, 1, ..., \dy}
    {
        \def\bit{0}
            \pgfmathsetmacro{\bit}{mod((\dx+\y),2)}
        \path [fill=lrcolor\bit] (\dx+1+\sftrux,\y+\sftruy) to [out=0,in=-90] (\dx+1+0.5+\sftrux, \y+0.5+\sftruy) to [out=90,in=0] (\dx+1+\sftrux,\y+1+\sftruy) --(\dx+1+\sftrux,\y+\sftruy);
        \path [fill=lrcolor\bit] (0+\sftrux,\y+1+\sftruy) to [out=180,in=-90] (-0.5+\sftrux, \y+1.5+\sftruy) to [out=90,in=180] (0+\sftrux,\y+2+\sftruy) --(0+\sftrux,\y+1+\sftruy);
    }
\foreach \x in {0, ..., 5}
    \foreach \y in {0, ..., 5}
    {
        \draw [black!60!,fill=black!60!] (\x,\y) circle [radius=0.1];

    }
\foreach \x in {-1, ..., 5}
    \foreach \y in {-1, ..., 5}
    {
        \draw [] (\x+0.5,\y+0.5) circle [radius=0.1];

    }

\node [white] (ctrl) at (1,1) {\fontsize{30}{30} \selectfont{C}};
\node [white] (ctrl) at (1,4) {\fontsize{30}{30} \selectfont{A}};
\node [white] (ctrl) at (4,4) {\fontsize{30}{30} \selectfont{T}};

\fill[white] (2.8, -0.7) rectangle (5.7, 2.2);
\fill[colorx] (3.3, 1.3) rectangle (3.6, 1.6);
\fill[colorz] (3.3, 0.5) rectangle (3.6, 0.8);
\node [colorx] (ctrl) at (4.5,1.45) {\fontsize{15}{15} \selectfont{X}};
\node [colorz] (ctrl) at (4.5,0.65) {\fontsize{15}{15} \selectfont{Z}};





\node (d1) at (0.2,0.1) {\fontsize{10}{10} \selectfont{0}};
\node (d1) at (1.2,0.1) {\fontsize{10}{10} \selectfont{1}};
\node (d1) at (2.2,0.1) {\fontsize{10}{10} \selectfont{2}};

\node (d1) at (0.2,1.1) {\fontsize{10}{10} \selectfont{3}};
\node (d1) at (1.2,1.1) {\fontsize{10}{10} \selectfont{4}};
\node (d1) at (2.2,1.1) {\fontsize{10}{10} \selectfont{5}};

\node (d1) at (0.2,2.1) {\fontsize{10}{10} \selectfont{6}};
\node (d1) at (1.2,2.1) {\fontsize{10}{10} \selectfont{7}};
\node (d1) at (2.2,2.1) {\fontsize{10}{10} \selectfont{8}};

\node (d1) at (0.2,3.1) {\fontsize{10}{10} \selectfont{9}};
\node (d1) at (1.2,3.1) {\fontsize{10}{10} \selectfont{10}};
\node (d1) at (2.2,3.1) {\fontsize{10}{10} \selectfont{11}};

\node (d1) at (0.2,4.1) {\fontsize{10}{10} \selectfont{12}};
\node (d1) at (1.2,4.1) {\fontsize{10}{10} \selectfont{13}};
\node (d1) at (2.2,4.1) {\fontsize{10}{10} \selectfont{14}};

\node (d1) at (0.2,5.1) {\fontsize{10}{10} \selectfont{15}};
\node (d1) at (1.2,5.1) {\fontsize{10}{10} \selectfont{16}};
\node (d1) at (2.2,5.1) {\fontsize{10}{10} \selectfont{17}};

\end{tikzpicture}\label{layout_cnot}}\hspace{5mm}
\subfigure[]{\begin{tikzpicture}

\colorlet{colorx}{blue!40!}
\colorlet{colorz}{red!40!}

\colorlet{color0.0}{colorx}
\colorlet{color1.0}{colorz}
\colorlet{udcolor0.0}{colorx}
\colorlet{udcolor1.0}{white}
\colorlet{lrcolor0.0}{white}
\colorlet{lrcolor1.0}{colorz}

\def\dx{0}
    \pgfmathsetmacro{\dx}{1}

\def\dy{0}
    \pgfmathsetmacro{\dy}{4}

\foreach \x in {0, ..., \dx}
    \foreach \y in {0, ..., \dy}
    {
        \def\bit{0}
            \pgfmathsetmacro{\bit}{mod((\x+\y+1),2)}
            \fill[color\bit] (\x, \y) rectangle (\x+1, \y+1);
    
    }

\foreach \x in {0, ..., \dx}
    {
        \def\bit{0}
            \pgfmathsetmacro{\bit}{mod((\x+\dy),2)}
        \path [fill=udcolor\bit] (\x,\dy+1) to [out=90,in=180] (\x+0.5, \dy+1.5) to [out=0,in=90] (\x+1,\dy+1) --(\x,\dy+1);
         \path [fill=udcolor\bit] (\x,0) to [out=-90,in=180] (\x+0.5, -0.5) to [out=0,in=-90] (\x+1,0) --(\x,0);
       
    }

\foreach \y in {0, ..., \dy}
    {
        \def\bit{0}
            \pgfmathsetmacro{\bit}{mod((\dx+\y),2)}
        \path [fill=lrcolor\bit] (\dx+1,\y) to [out=0,in=-90] (\dx+1+0.5, \y+0.5) to [out=90,in=0] (\dx+1,\y+1) --(\dx+1,\y);
       
    }
    
\foreach \y in {0, ..., \dy}
    {
        \def\bit{0}
            \pgfmathsetmacro{\bit}{mod((\dx+\y+1),2)}
        \path [fill=lrcolor\bit] (0,\y) to [out=180,in=-90] (-0.5, \y+0.5) to [out=90,in=180] (0,\y+1) --(0,\y);
       
    }

\foreach \x in {0, ..., 2}
    \foreach \y in {0, ..., 5}
    {
        \draw [black!60!,fill=black!60!] (\x,\y) circle [radius=0.1];
       
    }
\foreach \x in {-1, ..., 2}
    \foreach \y in {-1, ..., 5}
    {
        \draw [] (\x+0.5,\y+0.5) circle [radius=0.1];

    }

\node [white] (ctrl) at (1,1) {\fontsize{30}{30} \selectfont{C}};
\node [white] (ctrl) at (1,4) {\fontsize{30}{30} \selectfont{A}};
\draw [white,fill=white] (1,-0.7) circle [radius=0.1];

\node (d1) at (0.2,0.1) {\fontsize{10}{10} \selectfont{0}};
\node (d1) at (1.2,0.1) {\fontsize{10}{10} \selectfont{1}};
\node (d1) at (2.2,0.1) {\fontsize{10}{10} \selectfont{2}};

\node (d1) at (0.2,1.1) {\fontsize{10}{10} \selectfont{3}};
\node (d1) at (1.2,1.1) {\fontsize{10}{10} \selectfont{4}};
\node (d1) at (2.2,1.1) {\fontsize{10}{10} \selectfont{5}};

\node (d1) at (0.2,2.1) {\fontsize{10}{10} \selectfont{6}};
\node (d1) at (1.2,2.1) {\fontsize{10}{10} \selectfont{7}};
\node (d1) at (2.2,2.1) {\fontsize{10}{10} \selectfont{8}};

\node (d1) at (0.2,3.1) {\fontsize{10}{10} \selectfont{9}};
\node (d1) at (1.2,3.1) {\fontsize{10}{10} \selectfont{10}};
\node (d1) at (2.2,3.1) {\fontsize{10}{10} \selectfont{11}};

\node (d1) at (0.2,4.1) {\fontsize{10}{10} \selectfont{12}};
\node (d1) at (1.2,4.1) {\fontsize{10}{10} \selectfont{13}};
\node (d1) at (2.2,4.1) {\fontsize{10}{10} \selectfont{14}};

\node (d1) at (0.2,5.1) {\fontsize{10}{10} \selectfont{15}};
\node (d1) at (1.2,5.1) {\fontsize{10}{10} \selectfont{16}};
\node (d1) at (2.2,5.1) {\fontsize{10}{10} \selectfont{17}};

\end{tikzpicture}\label{merge}}\hspace{3.5mm}
\subfigure[]{\begin{tikzpicture}

\colorlet{colorx}{blue!40!}
\colorlet{colorz}{red!40!}

\colorlet{color0.0}{colorx}
\colorlet{color1.0}{colorz}
\colorlet{udcolor0.0}{colorx}
\colorlet{udcolor1.0}{white}
\colorlet{lrcolor0.0}{white}
\colorlet{lrcolor1.0}{colorz}

\def\dx{0}
    \pgfmathsetmacro{\dx}{1}
\def\dy{0}
    \pgfmathsetmacro{\dy}{1}

\def\Dx{0}
    \pgfmathsetmacro{\Dx}{\dx+1}
\def\Dy{0}
    \pgfmathsetmacro{\Dy}{\dy+1}
    
\def\sftonex{0}
    \pgfmathsetmacro{\sftonex}{0}
\def\sftoney{0}
    \pgfmathsetmacro{\sftoney}{3}
    
\def\sfttwox{0}
    \pgfmathsetmacro{\sfttwox}{3}
\def\sfttwoy{0}
    \pgfmathsetmacro{\sfttwoy}{0}
    
\def\sftldx{0}
    \pgfmathsetmacro{\sftldx}{0}
\def\sftldy{0}
    \pgfmathsetmacro{\sftldy}{0}
    
\def\sftrux{0}
    \pgfmathsetmacro{\sftrux}{3}
\def\sftruy{0}
    \pgfmathsetmacro{\sftruy}{3}

    
    
    
    
\foreach \x in {0,..., \dx}
    \foreach \y in {0, ..., \dy}
    {
        \def\bit{0}
            \pgfmathsetmacro{\bit}{mod((\x+\y),2)}
            \fill[color\bit] (\x+\sftonex, \y+\sftoney) rectangle (\x+1+\sftonex, \y+1+\sftoney);
    
    }

\foreach \x in {0,..., \dx}
    {
        \def\bit{0}
            \pgfmathsetmacro{\bit}{mod((\x+\dy+1),2)}
        \path [fill=udcolor\bit] (\x+\sftonex,\dy+1+\sftoney) to [out=90,in=180] (\x+0.5+\sftonex, \dy+1.5+\sftoney) to [out=0,in=90] (\x+1+\sftonex,\dy+1+\sftoney) --(\x+\sftonex,\dy+1+\sftoney);
        \path [fill=udcolor\bit] (\x+1+\sftonex,0+\sftoney) to [out=-90,in=180] (\x+1.5+\sftonex, -0.5+\sftoney) to [out=0,in=-90] (\x+2+\sftonex,0+\sftoney) --(\x+1+\sftonex,0+\sftoney);
    }

\foreach \y in {0, ..., \dy}
    {
        \def\bit{0}
            \pgfmathsetmacro{\bit}{mod((\dx+1+\y),2)}
        \path [fill=lrcolor\bit] (\dx+1+\sftonex,\y+\sftoney) to [out=0,in=-90] (\dx+1+0.5+\sftonex, \y+0.5+\sftoney) to [out=90,in=0] (\dx+1+\sftonex,\y+1+\sftoney) --(\dx+1+\sftonex,\y+\sftoney);
        \path [fill=lrcolor\bit] (0+\sftonex,\y-1+\sftoney) to [out=180,in=-90] (-0.5+\sftonex, \y-1+0.5+\sftoney) to [out=90,in=180] (0+\sftonex,\y-1+1+\sftoney) --(0+\sftonex,\y-1+\sftoney);
    }

\foreach \x in {0, 1, ..., \dx}
    \foreach \y in {0, 1, ..., \dy}
    {
        \def\bit{0}
            \pgfmathsetmacro{\bit}{mod((\x+\y+1),2)}
            \fill[color\bit] (\x, \y) rectangle (\x+1, \y+1);
    
    }

\foreach \x in {0, 1, ..., \dx}
    {
        \def\bit{0}
            \pgfmathsetmacro{\bit}{mod((\x+\dy),2)}
        \path [fill=udcolor\bit] (\x,\dy+1) to [out=90,in=180] (\x+0.5, \dy+1.5) to [out=0,in=90] (\x+1,\dy+1) --(\x,\dy+1);
        \path [fill=udcolor\bit] (\x-1,0) to [out=-90,in=180] (\x+0.5-1, -0.5) to [out=0,in=-90] (\x+1-1,0) --(\x-1,0);
    }

\foreach \y in {0, 1, ..., \dy}
    {
        \def\bit{0}
            \pgfmathsetmacro{\bit}{mod((\dx+\y),2)}
        \path [fill=lrcolor\bit] (\dx+1,\y) to [out=0,in=-90] (\dx+1+0.5, \y+0.5) to [out=90,in=0] (\dx+1,\y+1) --(\dx+1,\y);
        \path [fill=lrcolor\bit] (0,\y+1) to [out=180,in=-90] (-0.5, \y+1.5) to [out=90,in=180] (0,\y+2) --(0,\y+1);
    }



\foreach \x in {0, ..., 2}
    \foreach \y in {0, ..., 5}
    {
        \draw [black!60!,fill=black!60!] (\x,\y) circle [radius=0.1];

    }
\foreach \x in {-1, ..., 2}
    \foreach \y in {-1, ..., 5}
    {
        \draw [] (\x+0.5,\y+0.5) circle [radius=0.1];

    }

\node [white] (ctrl) at (1,1) {\fontsize{30}{30} \selectfont{C}};
\node [white] (ctrl) at (1,4) {\fontsize{30}{30} \selectfont{A}};

\draw [white,fill=white] (1,-0.7) circle [radius=0.1];

\node (d1) at (0.2,0.1) {\fontsize{10}{10} \selectfont{0}};
\node (d1) at (1.2,0.1) {\fontsize{10}{10} \selectfont{1}};
\node (d1) at (2.2,0.1) {\fontsize{10}{10} \selectfont{2}};

\node (d1) at (0.2,1.1) {\fontsize{10}{10} \selectfont{3}};
\node (d1) at (1.2,1.1) {\fontsize{10}{10} \selectfont{4}};
\node (d1) at (2.2,1.1) {\fontsize{10}{10} \selectfont{5}};

\node (d1) at (0.2,2.1) {\fontsize{10}{10} \selectfont{6}};
\node (d1) at (1.2,2.1) {\fontsize{10}{10} \selectfont{7}};
\node (d1) at (2.2,2.1) {\fontsize{10}{10} \selectfont{8}};

\node (d1) at (0.2,3.1) {\fontsize{10}{10} \selectfont{9}};
\node (d1) at (1.2,3.1) {\fontsize{10}{10} \selectfont{10}};
\node (d1) at (2.2,3.1) {\fontsize{10}{10} \selectfont{11}};

\node (d1) at (0.2,4.1) {\fontsize{10}{10} \selectfont{12}};
\node (d1) at (1.2,4.1) {\fontsize{10}{10} \selectfont{13}};
\node (d1) at (2.2,4.1) {\fontsize{10}{10} \selectfont{14}};

\node (d1) at (0.2,5.1) {\fontsize{10}{10} \selectfont{15}};
\node (d1) at (1.2,5.1) {\fontsize{10}{10} \selectfont{16}};
\node (d1) at (2.2,5.1) {\fontsize{10}{10} \selectfont{17}};

\end{tikzpicture}\label{split}}}
\caption{ (a)Three planar SC-based logical qubits with $d=3$. 
A 90-degree elbow-shaped qubit layout is required for implementing a lattice surgery-based $\cnot$ gate between qubits `C' and `T'. 
`C' is the control qubit, `T' is the target qubit, and `A' is the ancilla qubit in either $\ket{0}$ or $\ket{+}$ state. 
(b) The integrated lattice `AC' and (c) the separated lattices `A' and `C', after merging and splitting `A' and `C', respectively.}
\label{cnot}
\end{figure}
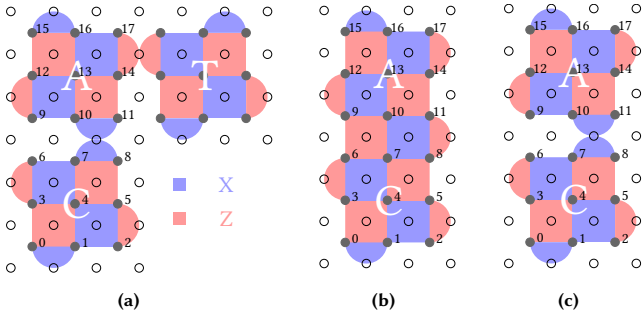

\subsection{Fault-tolerant mechanisms}
Figure~\ref{layout_cnot} shows three logical qubits based on distance-$3$ planar SC and they are labeled as `A', `T' and `C', respectively. 
Each logical qubit consists of $17$ physical qubits and has two types of boundaries,  $Z$-boundaries and $X$-boundaries. For instance, in lattice `A', the left and right boundaries are $Z$-type and the top and bottom boundaries are $X$-type.
In planar SC, initialization, measurement, Pauli gates, and $H$ can be implemented transversally, i.e., applying bitwise physical operations on a subset of the data qubits, and then performing ESM to detect errors. 
The FT implementation of $S$ and $T$ gates in surface code requires ancillary qubits prepared in specific states called magic states. 
However, the preparation of magic states is not fault-tolerant and produces states with low fidelity that need to be purified by a non-deterministic procedure called state distillation \cite{bravyi2005universal, bravyi2012magic, meier2012magic, jones2013multilevel, campbell2017unifying}. 
This distillation procedure is repeated until the measurement results indicate one state is successfully purified. On top of that, multiple rounds of successful distillation may be required to achieve the desired state fidelity.
Therefore, magic state distillation is the most resource- and time-consuming process in FT quantum computing. 
Since the $S$ and $T$ gates can be performed only if their corresponding magic states have been delivered, an online or dynamic scheduling and run-time routing may be required for efficient circuit execution~\cite{paler2017online}. 
In this paper, we assume magic states have been prepared and properly allocated whenever $S$ and $T$ gates need to be performed. 
We will investigate the dynamics of magic state preparation in future work.

\begin{figure}[tb!]
\centering
\subfigure[]{\includegraphics[width=1.6in]{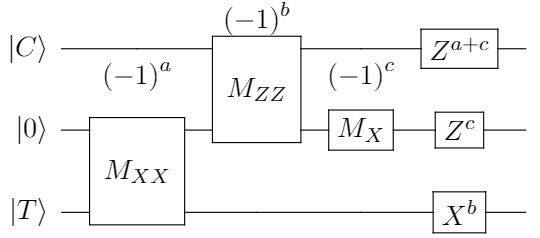}
\label{cnot_circuit1}}
\subfigure[]{\includegraphics[width=1.6in]{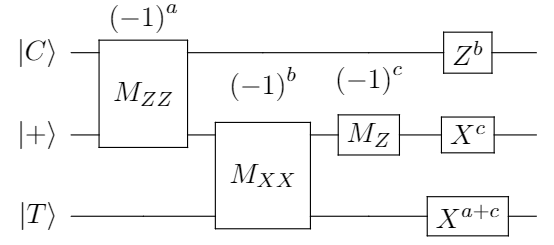}
\label{cnot_circuit2}}
\caption{The circuits (a) and (b) to realize the measurement-based $\cnot$ gates. 
`$C$' is the control qubit, `$T$' is the target qubit. An ancillary qubit is prepared in state $\ket{0}_{L}$ (a) or $\ket{+}_{L}$ (b). }
\label{cnot_circuit}
\end{figure}

In principle, a FT logical $\cnot$ gate between two planar logical qubits can be performed transversally, i.e., applying pairwise physical $\cnot$ gates to the data qubits in the two lattices. 
However, this transversal $\cnot$ cannot be realized in current quantum technologies which only allow NN interactions in 2D architectures. 
Alternatively, a measurement-based procedure~\cite{gottesman1998fault} which is equivalent to a $\cnot$ gate can be applied and its circuit representations are shown in Figure~\ref{cnot_circuit}. 
The joint measurement $M_{XX}$ ($M_{ZZ}$) is realized by first merging two logical qubits and then splitting them, where their adjacent boundaries are $Z$-($X$-)type boundaries. 
The outcomes of these measurements will determine whether the corresponding Pauli corrections should be applied (see Appendix ~\ref{app:cnot} for more details).

The qubit layout for performing the measurement-based $\cnot$ gate in the 2D NN architecture is shown in Figure \ref{layout_cnot}.
The realization of the circuit in Figure~\ref{cnot_circuit2} is achieved as follows: 1) lattices `A' and `C' are merged and then split; 2) lattices `A' and `T' are merged and then split; and 3) measure `A'.
The merge and split operations are implemented by a technique called lattice surgery \cite{horsman2012surface, landahl2014quantum}. 
For instance, the merge and split of lattice `A' and `C' are implemented by performing ESM on the integrated lattice (Figure~\ref{merge}) and on the separated lattices (Figure~\ref{split}), respectively. 
In general, a surgery-based CNOT takes $4d+1$ SC cycles. It is worthy to mention that a split operation between qubits `$A$' and `$C$'(`$T$') can happen simultaneously with a merge operation between qubits `$A$' and `$T$'(`$C$'). Furthermore, a split operation between two qubits and a measurement on one of them can be performed in parallel. By exploiting the parallelism, the execution time in SC cycles can be reduced to $3d$.


\subsection{Implications on the mapping problem}
Based on the FT implementation of logical operations on planar SC, we derive the following constraints that must be taken into account by the mapping process as well as its implications. 

\textbf{Constraints:} 1)
The physical 2D NN interaction constraint is intrinsically satisfied by the construction of surface codes, thus the physical-level mapping becomes trivial; 
2) A surgery-based CNOT gate requires that the qubits `C' and `T' together with the ancilla qubit `A' are placed in particular neighbouring positions, forming a 90-degree elbow-shaped layout.

\textbf{Implications:} 1) Logical qubits that need to interact and are not placed in such neighbouring positions need to be moved, for instance by means of SWAP operations. The movement of qubits introduces overhead in terms of both qubit resources and execution time; 
2) Therefore, in lattice surgery-based SC quantum computing, it is essential to pre-define a qubit plane architecture for efficiently managing qubit resources and supporting communication between logical qubits;
3) In addition, operations for moving qubits should be defined;
4) It is necessary to initially place highly interacting logical qubits as close as possible and apply routing techniques to find the communication paths.

Based on the above observations, we will introduce two slightly different plane architectures and mapping passes for efficient execution of lattice surgery-based quantum circuits in the following sections.

\section{Qubit plane architecture} \label{sec-arch}

\begin{figure}[bt!]
\centering
\includegraphics[width=1.5in]{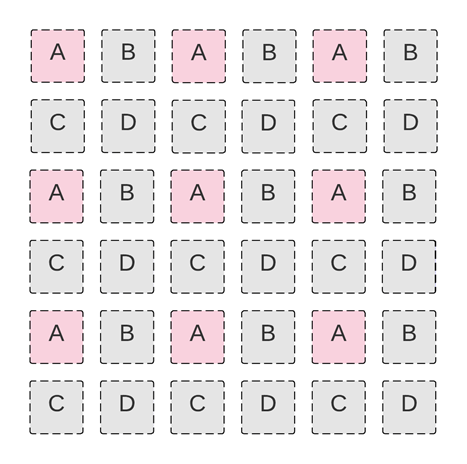}
\caption{The qubit plane architecture proposed in \cite{horsman2012surface} for lattice surgery-based planar surface codes, where each patch can hold one logical qubit shown in Figure~\ref{layout_cnot}.}
\label{basiclayout}
\end{figure}

\begin{figure}[bt!]
\centerline{\subfigure[]{\includegraphics[width=1.5in]{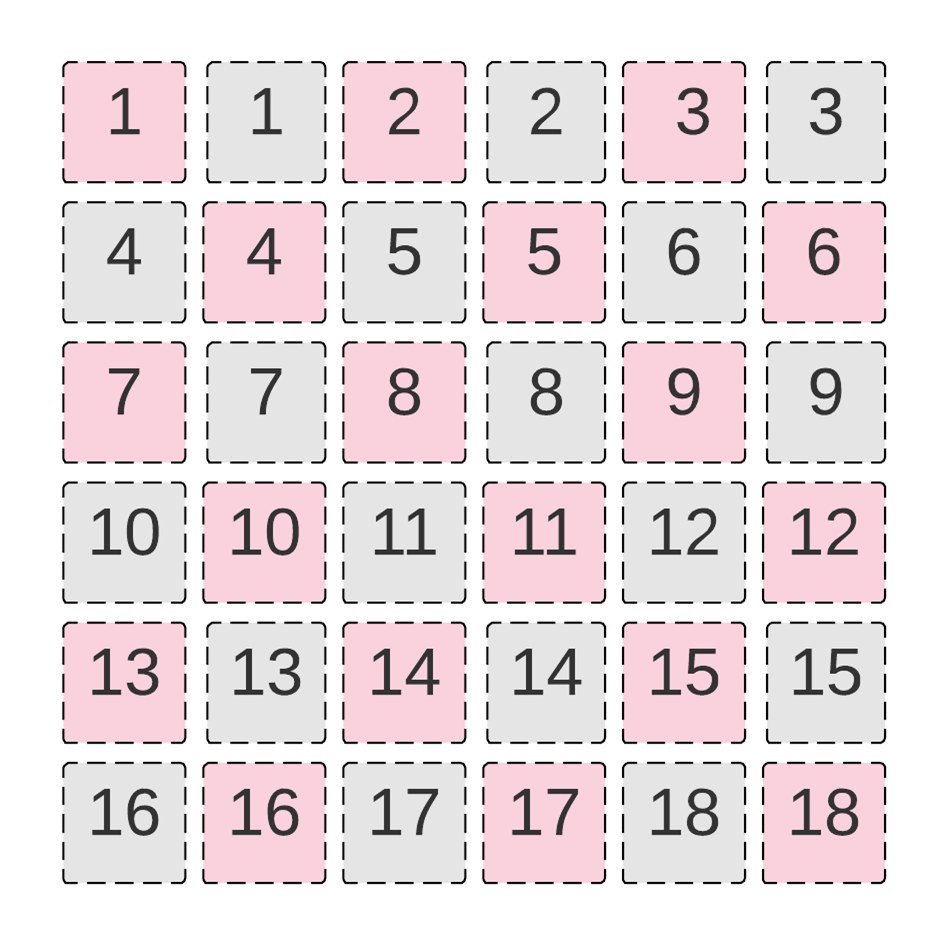}\label{checker}}\hspace{5mm}\subfigure[]{\includegraphics[width=1.5in]{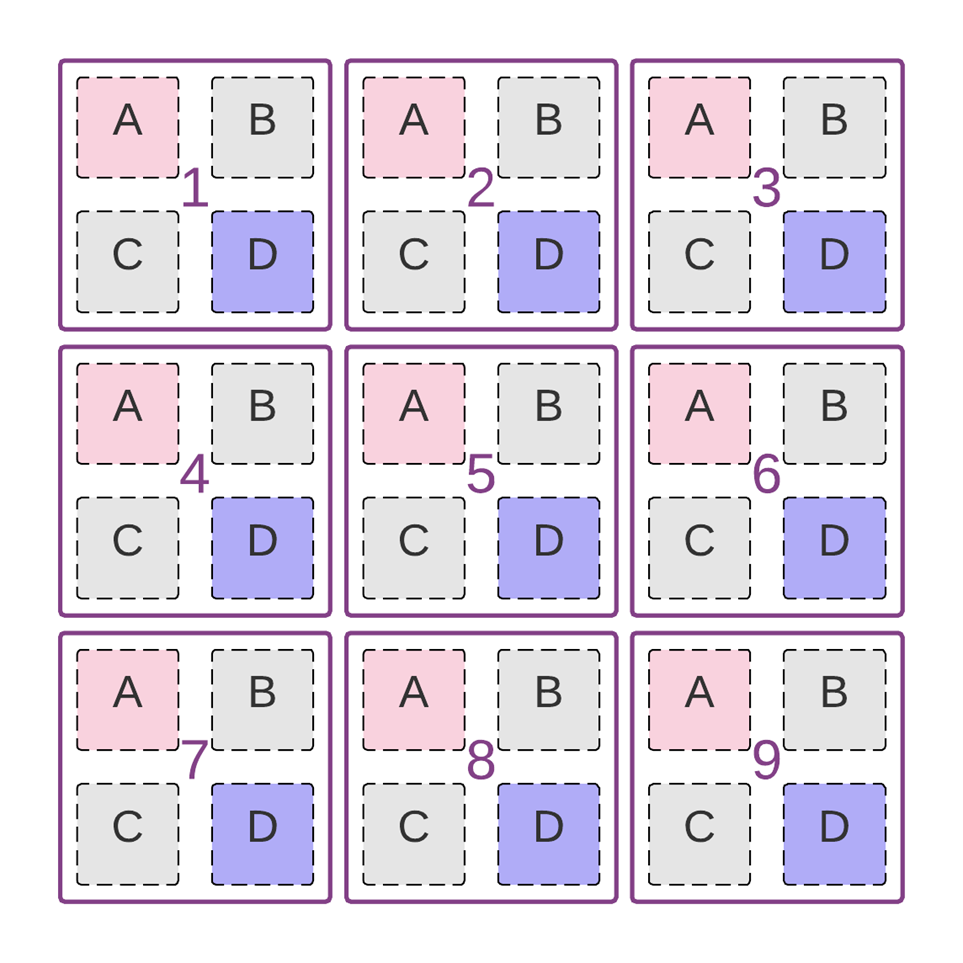}\label{tile}}}\vspace{-5mm}
\caption{(a) The checkerboard architecture (c-arch)  and (b) the tile-based architecture (t-arch). }
\label{planar_layout}
\end{figure}

A qubit plane architecture is a virtual layer that organizes the qubits in different specialized and pre-defined areas such as communication, computation and storage \cite{dousti2014squash, heckey2015compiler}. Qubit architectures should be able to manage qubit resources efficiently and provide fast execution of any quantum circuit.

In ~\cite{horsman2012surface}, a layout that supports lattice surgery-based $\cnot$ gates on planar SC is presented. As shown in Figure \ref{basiclayout}, it consists of several patches. The gray patches of the lattice are used for allowing qubits to perform CNOT operations,  whereas the pink patches are used for holding logical data qubits. Then, only $1/4$ of the available patches contains logical data qubits. Based on this layout, we propose two slightly different  qubit plane architectures, the \textit{tile-based} architecture (t-arch) and the \textit{checkerboard} architecture (c-arch) as shown in Figure~\ref{planar_layout}. The pink and purple patches are where logical data qubits containing information can be allocated (data patches), whereas the gray patches are assisting logical qubits (ancilla patches) that are used for performing logical $\cnot$ gates and for communication. 
These two architectures differ in: i) the number of logical data qubits that can allocate, ii) the way movement operations are implemented, iii) the steps required for performing a CNOT between neighbouring logical data qubits, and iv) the number of neighbours. 



\textbf{Logical data qubit allocation:} 
In the checkerboard architecture, logical data qubits can be assigned to any of the pink patches, that is, $1/2$ of the total patches are used to hold data qubits.
In the tile-based architecture, a lined area consisting of $4$ logical patches is defined as a basic computation tile and at most one logical data qubit can be allocated in each tile, that is, in either the pink or the purple patch. 
Then, only $1/4$ of the total number of patches can be used for allocating logical data qubits.


\textbf{Movement operations:}
One typical way to move physical qubits is through SWAP operations in which the state of the qubits is exchanged. Usually, a SWAP gate is implemented by applying $3$ consecutive CNOT gates. 
The same principle can be applied for moving logical qubits. 
In this case a logical SWAP is realized by performing $3$ consecutive lattice surgery-based $\cnot$ gates, which is extremely time-consuming ($9d$ SC cycles).
In the checkerboard architecture, we will use such a swap method called \textbf{c-SWAP} for moving logical qubits because of the limited number of ancilla patches. 
In the tile-based architecture, we propose to use a faster movement operation, which is analogous to the measurement-based procedure for $\cnot$ gates, to swap data information between two horizontally or vertically adjacent tiles. 
This swap operation called \textbf{t-SWAP} only takes $1$x logical $\cnot$ gate time regardless of locations where data qubits are allocated inside the tiles -i.e. purple or pink patches. 
It is realized by 'moving' qubits to neighbouring horizontal and vertical patches (see Appendix~\ref{app:movement}).
Figure~\ref{tswap} shows an example of how to swap two logical data qubits placed in adjacent tiles by using the t-SWAP operation. Similarly, one can perform a t-SWAP between any other pair of patches in the horizontally or vertically adjacent tiles. 



\begin{figure}[tb]
\centerline{\subfigure[]{\includegraphics[width=0.9in]{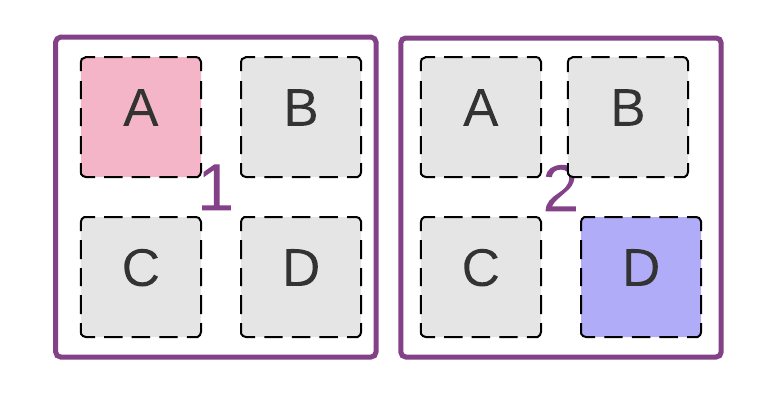}
\label{tswap1}}
\subfigure[]{\includegraphics[width=0.9in]{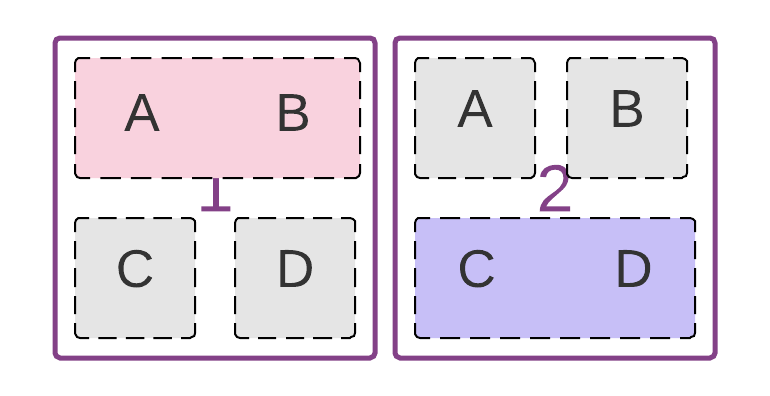}
\label{tswap2}}
\subfigure[]{\includegraphics[width=0.9in]{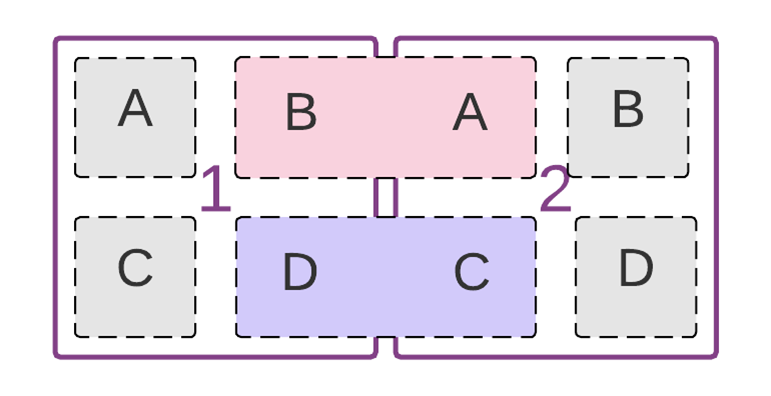}
\label{tswap3}}
\subfigure[]{\includegraphics[width=0.9in]{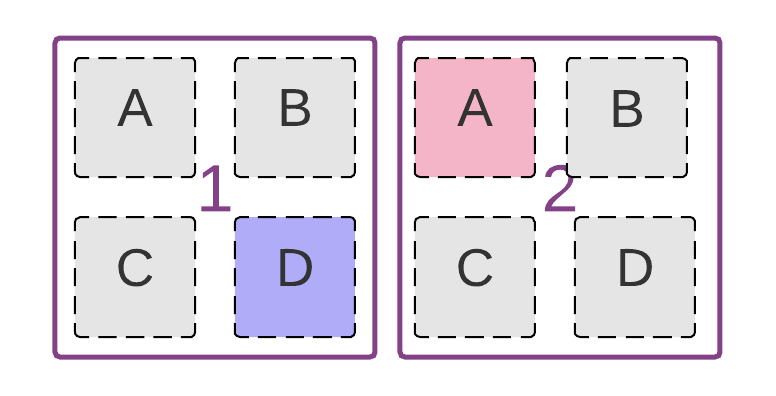}
\label{tswap4}}}
\caption{A t-SWAP between tiles 1 and 2. (a) Logical qubits are in patches A1 and D2. (b) Merge A1, B1 and D2, C2; (c) merge B1, A2 and C2,D1, and measure D2, A1; (d) measure B1, C2.}
\label{tswap}
\end{figure}

\begin{figure}[tb]
\centerline{\subfigure[]{\includegraphics[width=0.9in]{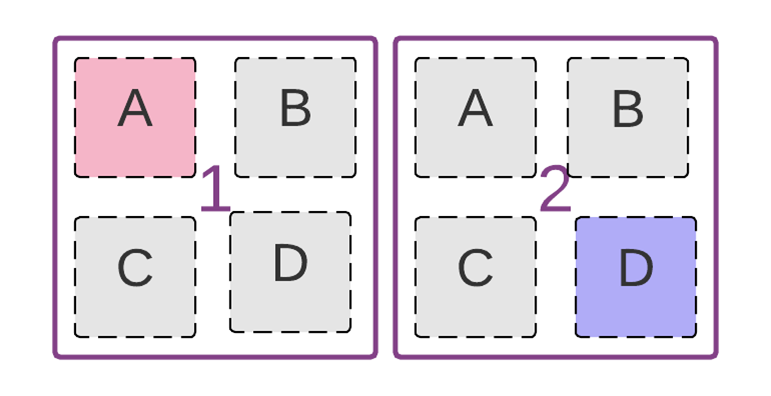}
\label{tcnot1}}
\subfigure[]{\includegraphics[width=0.9in]{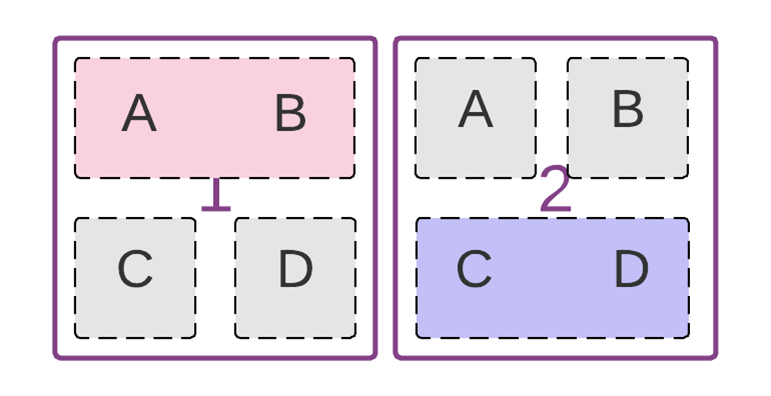}
\label{tcnot2}}
\subfigure[]{\includegraphics[width=0.9in]{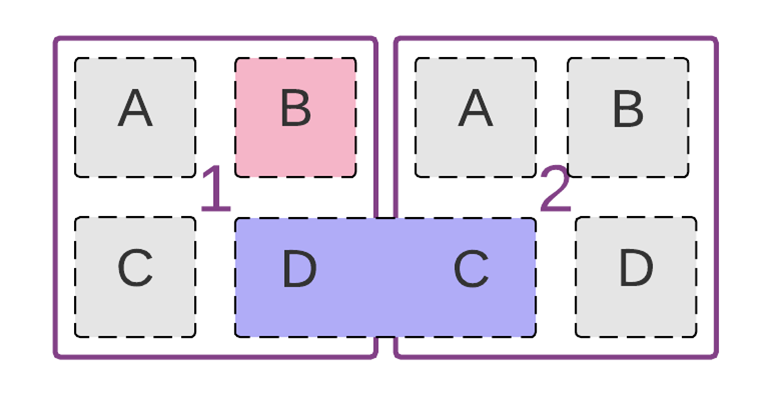}
\label{tcnot3}}}
\centerline{\subfigure[]{\includegraphics[width=0.9in]{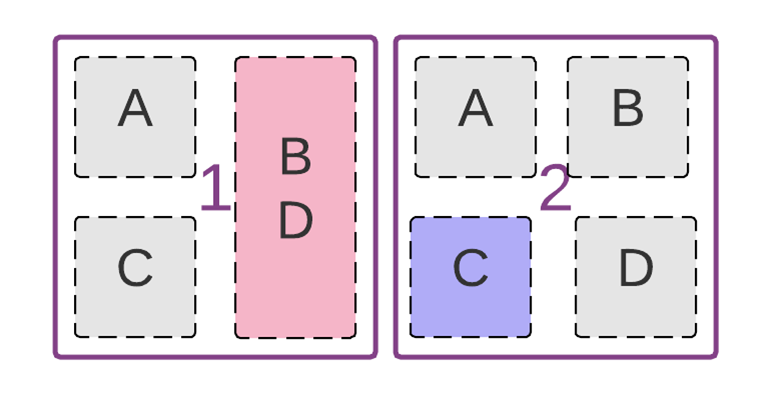}
\label{tcnot4}}
\subfigure[]{\includegraphics[width=0.9in]{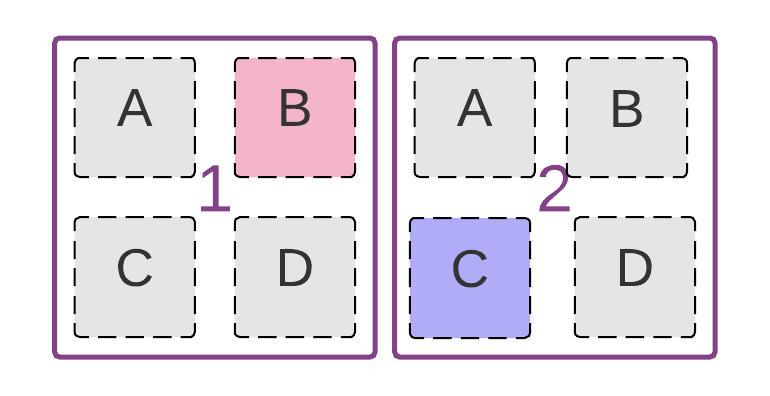}
\label{tcnot5}}}
\caption{A t-CNOT between tiles 1 and 2. (a) Control and target qubits are in patches A1 and D2 (b) Merge A1, B1 and D2, C2; (c) measure A1 and D2 and merge D1, C2; (d) merge B1,D1; (e) measure D1. The CNOT is performed in steps c), d) and e). }
\label{tcnot}
\end{figure}

\textbf{CNOT operations:}
As mentioned in Section~\ref{sec_ft}, the control and target qubits need to be placed in patches that form a 90-degree elbow-shaped in order to perform a lattice surgery-based CNOT. 
In the checkerboard architecture, two neighbouring data patches are always in such 90-degree locations so that a lattice surgery-based $\cnot$ gate can be directly performed between them. 
We called this operation $\textbf{c-CNOT}$ and it is implemented by 3 steps, taking $3d$ SC cycles as described in the previous section. 
However, in the tile-based architecture, a $\cnot$ operation called $\textbf{t-CNOT}$ between two data qubits placed in horizontally, vertically, or diagonally adjacent tiles may need some pre-processing, depending on where data qubits are allocated. 
If the control and target logical qubits are already placed in patches forming a 90-degree shape, then one can perform the CNOT directly, e.g., patch D1 with patches A4, A2, A5. 
Otherwise, logical data qubits need to be moved to the required locations before performing the CNOT gate as shown in Figure \ref{tcnot}. 



Similarly, one can perform a t-CNOT between any other pair of patches in adjacent tiles. 
The t-CNOT with and without pre-processing takes $4d$ and $3d$ SC cycles, respectively. 
In the results section, we will assume that a t-CNOT always takes 4d SC cycles for simplicity.


\textbf{Number of neighbours:} 
In the checkerboard architecture one data patch can only interact with $4$ adjacent data patches, e.g., the neighbours of patch $8$ are $4,5,10,11$ in Figure \ref{checker}. 
As mentioned in Section~\ref{sec_ft}, a logical ancilla is required for performing a lattice surgery-based $\cnot$ gate. 
To avoid ancilla conflicts when performing multiple logical $\cnot$ gates simultaneously in the checkerboard, only the upper ancilla patch adjacent to the two interacting data patches can be used. 
For instance, ancilla $1$ ($2$) will be used when performing a $\cnot$ between data qubits $2$ and $4$ ($2$ and $5$).
In the tile-based architecture, one tile can interact with at most $8$ neighbours, e.g., the neighbours of tile $5$ are $1,2,3,4,6,7,8,9$. 
However, logical $\cnot$ gates between data qubits in tiles $1$ and $5$, and between data qubits in tiles $2$ and $4$ cannot be performed simultaneously because of ancilla conflicts. 
To avoid such conflicts for now, we only assume 6 neighbours per tile; we remove the right-top and left-bottom neighbours of each tile, e.g., remove tiles $3$ and $7$ from the neighbour list of tile $5$. 

In the next section, we will introduce the procedure for mapping lattice surgery-based quantum circuits onto both qubit architectures. 
We will then evaluate their communication overhead in terms of both operation overhead and latency overhead in Section~\ref{sec-result} .


\section{Quantum circuit mapping}
\label{sec-mapping}


The mapping of quantum circuits involves initial placement and routing of qubits and scheduling of operations. 
The need for QEC significantly enlarges the circuit size, which makes the mapping problem even more complex. 
For instance, in surface codes one logical qubit is encoded into $O(d^{2})$ physical qubits and one logical operation is implemented by $O(d^{3})$ physical operations, where $d$ is the code distance. 
Therefore, we propose to perform the mapping of quantum circuits before going to the physical implementation of logical qubits and operations. 
It means that each logical qubit is treated as one single unit, and each logical operation is regarded as one single instruction. 
Once the mapping is finished, logical operations need to be expanded into the corresponding physical operations. 
We use a library to translate each logical operation into pre-scheduled physical quantum operations (see Appendix~\ref{app:ftlib}). 
During the translation, the address of underlying physical qubits corresponding to a logical qubit can be retrieved by maintaining a q-symbol table ~\cite{fu2016heterogeneous}.

Table~\ref{time_log} depicts the execution time of different logical operations on planar surface codes expressed in SC cycles. 
It includes single-qubit operations as well as the two-qubit operations used in both qubit architectures presented previously.
The execution time of different operations is used in the scheduling and routing passes. 
Furthermore, we will use these numbers for calculating the overall circuit latency in Section 5.  

In order to illustrate the different steps in the mapping of quantum circuits, we will use the circuit in Figure \ref{steane_QASM} described by a quantum assembly language (QASM). 
This is the encoding circuit of the $7$-qubit Steane code $[[7,1,3]]$ and it can also be used to distill the magic states for $S$ gates \cite{fowler2012surface}. In this case, we assume each qubit is a logical qubit encoded by a distance-$7$ planar SC and each operation is a FT operation implemented by the techniques in Section~\ref{sec_ft} and Section~\ref{sec-arch}.

\begin{table}[tb!]
\centering
\small
\caption{The execution time in SC cycles of different logical operations, $d$ is the code distance.}\vspace{-2mm}
\label{time_log}
\begin{tabular}{c|c|c|c|c|c}
\hline
          &Init \& MSMT& Pauli  &$H$   & $S$ & $T$   \\ \hline
\# Cycles  &$1$ &$1$ & $4d$ & $14d$ &$17d$   \\ \hline
& c-CNOT\ &c-SWAP   & t-CNOT & t-SWAP  \\ \hline \# Cycles  & $3d$ & $9d$ & $4d$ &$3d$  \\ \hline
\end{tabular}
\end{table}

\subsection{Scheduling operations}
\label{sec-schedule}
The objective of the scheduling problem is to minimize the total execution time (circuit latency) of quantum algorithms meanwhile keeping the correctness of the program semantics. 
Similar to instruction scheduling in classical processors, the correctness can be achieved by respecting the data dependency~\cite{hennessy2011computer} between quantum operations. 
Analogous to classical computing, two kinds of data dependency can be defined for quantum computing: true dependency, which is the dependency between two single-qubit gates and between a single-qubit gate and a $\cnot$ gate, and name dependency, which is the dependency between two $\cnot$ gates which have the same control (or target) qubit. 

\begin{figure}[tb]
\centerline{\subfigure[]{\includegraphics[width=0.8in]{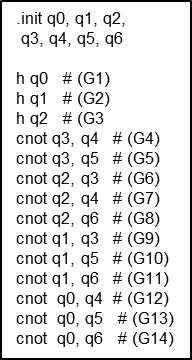}
\label{steane_QASM}}
\subfigure[]{\includegraphics[width=1.3in]{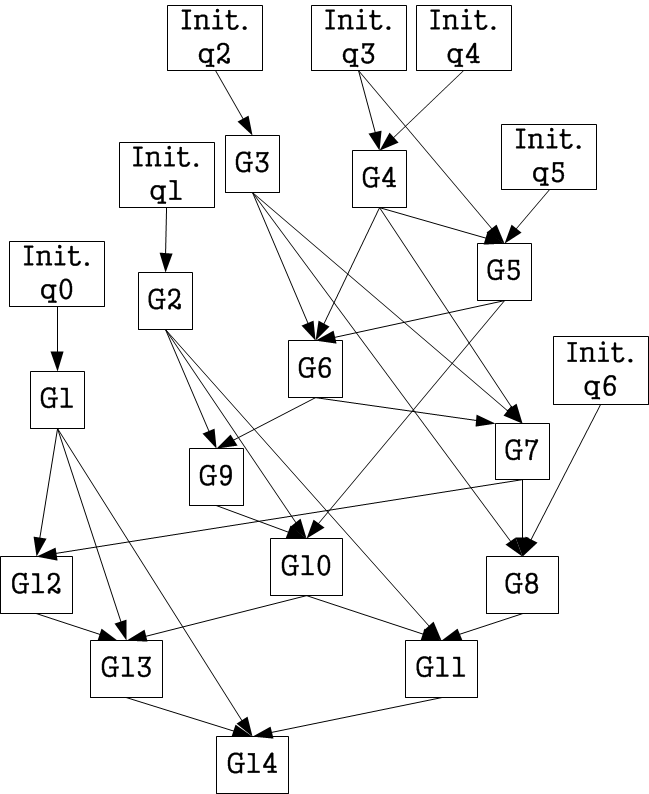}
\label{steane_QODG}}
\subfigure[]{\includegraphics[width=1.3in, height=1.5in]{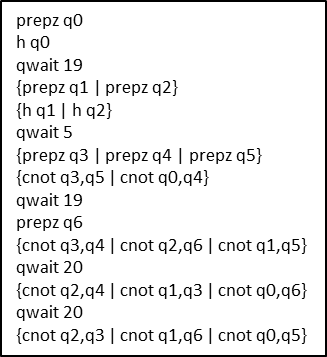}
\label{steane_qasm_para}}}
\caption{The QASM description of the Steane $[[7,1,3]]$ encoding circuit and its QODG. 
(a) The serial QASM representation; (b) The QODG; and (c) The scheduled parallel QASM representation. 
`qwait' is an instruction specifies the waiting time until the next instruction can be issued.}
\label{steane_qasm_qodg}
\end{figure}

We convert a QASM-described quantum circuit into a data flow-based weighted directed graph, which is called Quantum Operation Dependency Graph (QODG) and shown in Figure~\ref{steane_QODG}. 
In this graph $G(V_{G}, E_{G})$, each operation is denoted using a node $v_i$, and the data dependency arising from two consecutive operations on a same qubit, e.g., $v_i$ followed by $v_j$, is represented using a directed edge $e(v_i, v_j)$. $V_{G}$ and $E_{G}$ are the node set and edge set of $G$, respectively. We also define $E^1_G$ and $E^2_G$ as the collection of edges that exhibits true and name dependency, respectively. $S_{v_{i}}$ represents the starting time of operation $v_i$ and $T_{v_{i}}$ indicates its latency.
The scheduling objective is to minimize the total circuit latency (Formula \ref{equ:5}) while preserving the data dependency between operations (Formula \ref{equ:6}).
\begin{align}\label{equ:5}
\min \sup_{\forall v_{i} \in V_{G}}{(S_{v_{i}} + T_{v_{i}})}
\end{align}

\textit{subject to} \vspace{-3mm}
\begin{equation}\label{equ:6}
(S_{v_{i}} + T_{v_{i}}) \leqslant S_{v_{j}},\ \forall e(v_{i},v_{j}) \in E_{G}  
\end{equation}

Note that two $\cnot$ gates which share the same control or the same target qubit are commutable, meaning that they can be executed in any order except in parallel. 
This commutation property has not been considered in previous works~\cite{lin2015paqcs,metodi2006scheduling, whitney2007automated, dousti12min,yazdani2013quantum, bahreini2015minlp, balensiefer2005quale, dousti2013leqa, dousti2014squash, ahsan2015architecture, heckey2015compiler}. 
In this paper, we take commuted $\cnot$ gates into account and replace the optimization condition~\ref{equ:6} with conditions~\ref{equ:7} and \ref{equ:8}:
\begin{equation}\label{equ:7}
S_{v_{i}} + T_{v_{i}} \leqslant S_{v_{j}},\quad \forall e(v_{i},v_{j}) \in E^1_{G}  
\end{equation}
\begin{equation}\label{equ:8}
(S_{v_{i}} - S_{v_{j}}) \leqslant T_{v_{j}} \ or\ (S_{v_{j}} - S_{v_{i}}) \leqslant T_{v_{i}},\ \forall e(v_{i},v_{j}) \in E^2_{G}  
\end{equation}


With respect to different dependencies, the scheduler will exploit parallelism and output the operation sequence with timing information, which is an as-soon-as-possible (ASAP) schedule. 
An as-late-as-possible (ALAP) schedule can be also easily implemented by scheduling operations in the reverse order (Figure~\ref{steane_qasm_para}).

\subsection{Placing and routing qubits}
\begin{figure}[tb]
\centerline{\subfigure[]{\includegraphics[width=1in]{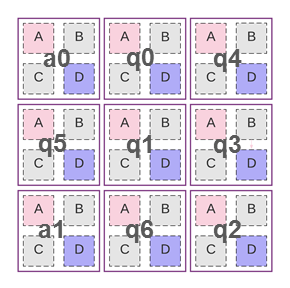}
\label{iplace_tarch_7enc}}
\subfigure[]{\includegraphics[width=1in]{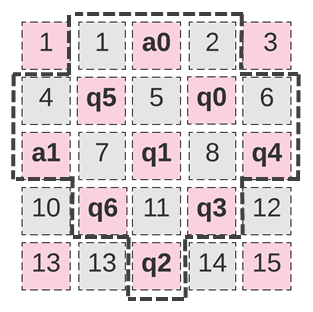}
\label{iplace_carch_7enc}}}
\caption{The initial placements of the Steane $[[7,1,3]]$ encoding circuit in (a) the tile-based architecture which has $3\times3$ tiles and (b) the checkerboard architecture which has $3\times3$ data patches in the dashed region (rotated by $45$ degrees).}
\label{iplace_7enc}
\end{figure}

{\bf The QAP-model for initial placement of qubits:}
The goal of qubit placement is to find an optimal initial placement of the qubits that minimizes communication overhead. 
Similar to the placement approaches in~\cite{bahreini2015minlp,dousti2014squash,shafaei2014qubit}, the initial placement problem is formulated as a quadratic assignment problem (QAP) with the communication overhead represented using the Manhattan distance: 
\begin{equation}\label{equ:1}
\min \left ( \sum_{i=1}^{m}\sum_{j=1}^{m}\sum_{k=1}^{n}\sum_{l=1}^{n}c_{ijkl}x_{ik}x_{jl} \right )
\end{equation}
  \textit{subject to}
\begin{equation}\label{equ:2}
\sum_{i=1}^{m}x_{ik}= 1, \quad \forall k = 1, \cdots n    
\end{equation}
\begin{equation}\label{equ:3}
\sum_{k=1}^{n}x_{ik}= 1, \quad \forall i = 1, \cdots m    
\end{equation}
\begin{equation}\label{equ:4}
x_{ik}= \left \{ 0,1 \right \}  
\end{equation}
where $m$($n$) is the number of locations(qubits), $x_{ik(jl)}=1$ or $0$ indicates whether qubit $k(l)$ is assigned to location $i(j)$ or not,  $c_{ijkl} = D_{ij} R_{kl}$ is the cost of separately assigning qubit $k$ and $l$ to locations $i$ and $j$. 
$D_{ij}$ is the Manhattan distance between locations $i$ and $j$, and $R_{kl}$ is the number of interactions between qubits $k$ and $l$ in the circuit. 
Constraints \ref{equ:2} and \ref{equ:3} ensure a one-to-one mapping from qubits to locations. 
A location is a tile in the tile-based architecture and a data patch in the checkerboard architecture.
For instance, the initial placements of the Steane $[[7,1,3]]$ encoding circuit in the $m=3\times 3$ tile-based architecture and the $m=3\times 3$ checkerboard architecture are shown in Figure \ref{iplace_7enc}.

In this paper, the scheduling and QAP models are solved with integer linear programming (ILP). 
The scheduling uses the linearization method by~\cite{richards2002spacecraft}, and the QAP uses the method proposed by~\cite{kaufman1978algorithm}.
ILP can only solve small-scale problems in reasonable time as the ones used in this paper. 
Even though for near-term implementation in FT quantum computing, these numbers largely suffice. 
For large-scale circuits, one can either partition a large circuit into several smaller ones or apply heuristic algorithms to efficiently solve these mapping models ~\cite{metodi2006scheduling, whitney2007automated, dousti12min, bahreini2015minlp,balensiefer2005quale, dousti2013leqa, ahsan2015architecture}. 

{\bf The routing algorithm:} The introduced two SC qubit architectures require routing of qubits, which involves finding communication paths and inserting the corresponding movement operations, for instance by means of the $\swap$ operations. 
An efficient routing should minimize the number of inserted movement operations as well as the increased latency. 
In this paper, qubits are routed based on a sliding window (buffer) principle as shown in Algorithm $1$. 
The algorithm will find a path for the first not NN instruction- i.e. CNOT operation in which qubits are not NN- inside the buffer. 
We adopted the breadth-first search (BFS) algorithm to find all possible shortest paths. 
Then, in order to select the communication path the algorithm looks back and forward. 
The look-back finds the maximum interleaving of movement instructions ($\swap$s) with previous instructions. 
The look-ahead will look how the positions of the qubits involved in a certain path is changed and how it affects future two-qubit operations; that is, we want to avoid to move away qubits that are already close to each other and need to interact in the future. 
Once the path is selected, the instructions inside the buffer will be rescheduled using the ASAP strategy. 
Then the buffer will output routing instructions and will be fed with new ones. 
This process repeats until all $\cnot$ gates can be performed in the pre-defined qubit architecture.

\def\NoNumber#1{{\def\alglinenumber##1{}\State #1}\addtocounter{ALG@line}{-1}}
\begin{algorithm}[tb!]
	\renewcommand{\algorithmicrequire}{\textbf{Input: }}
	\renewcommand{\algorithmicensure}{\textbf{Output: }}
	\caption{Routing algorithm}
	\begin{algorithmic}[1] 
		\Require \hspace{0.4em} Defined qubit architecture and its size,
		\par\hspace{0.3em} initial placement, scheduled QASM-file
		\Ensure Routed QASM-file 
		\State {Define instruction buffer, $B$, length $l$ = window size}
		\State Fill $B$ with instructions from input-QASM 
		\While {$B$ \textbf{is not} empty}
			\State{\color{gray} \# Check if an instruction ($ins$) is NN}
			\For {$ins$ \textbf{in} $B[0:l/2]$}
				\State{\color{gray} \# If an instruction is not NN, start routing}
				\If {$ins$ \textbf{is not} NN}
				    \State{\color{gray} \# Find different paths}
				    \State{$paths =$ all shortest paths for $ins$ based on BFS}
				    \State{\color{gray} \# Look-back}
				    \For {$p\ $\textbf{in} $paths$}
				        \State {$p.length\ =$ \#cycles from in $p$ $-$ \#cycles $p$ can}
				        \NoNumber {\hspace{0.4em}interleave with instruction in $B[0:ins]$}
				    \EndFor
				    \State{\color{gray} \# Look-ahead to other ins $(o\_i)$}
				    \For {$p\ $\textbf{in} $paths$}
				        \State {Place qubits based on $p$}
				        \State {$p.length \mathrel{+}= \sum\nolimits_{o\_i \in B[ins:l]}$ shortest path}
				        \NoNumber{\hspace{0.4em} for $o\_i$ in \#cycles}
				    \EndFor
				    \State{Insert path with min. length and update placement}
			        \State{\textbf{Break} for-loop}
			    \EndIf
			\EndFor
			\State{Reschedule $B[0:ins]$}
			\State{Write $B[0]$ to output-QASM}
			\State{Fill $B$ from input-QASM with qubit placement}
		\EndWhile
	\end{algorithmic}
\end{algorithm}

\begin{figure}[tb!]
\centering
\includegraphics[width=3in]{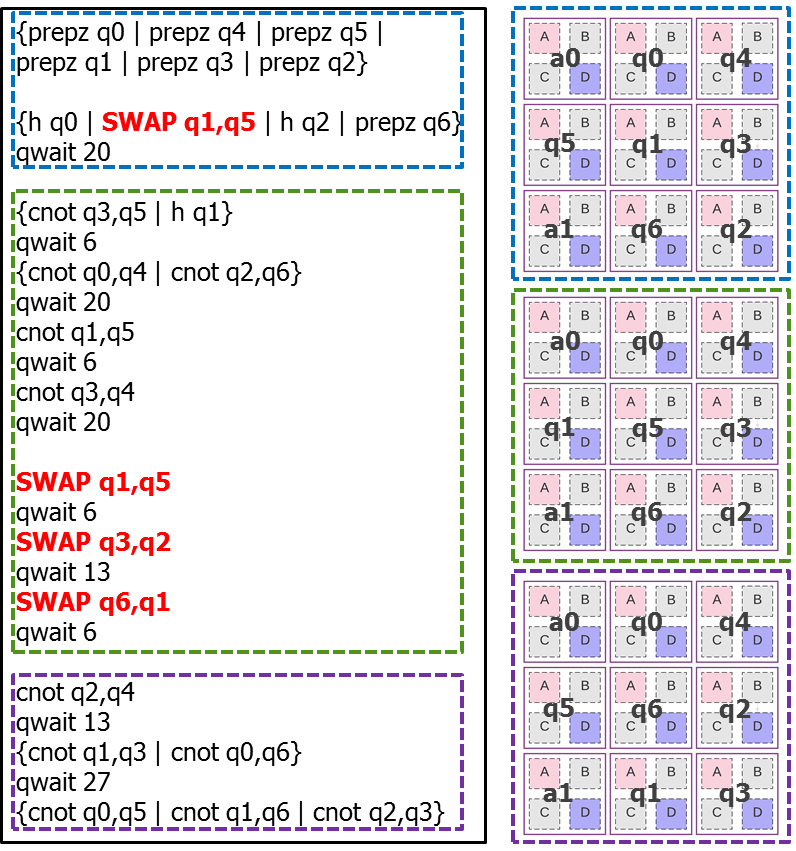}
\caption{The final circuit and the intermediate qubit layouts after mapping the Steane $[[7,1,3]]$ encoding circuit onto the tile-based architecture.}
\label{maptarch_7enc}
\end{figure}

\begin{figure}[tb!]
\centering
\includegraphics[width=3in]{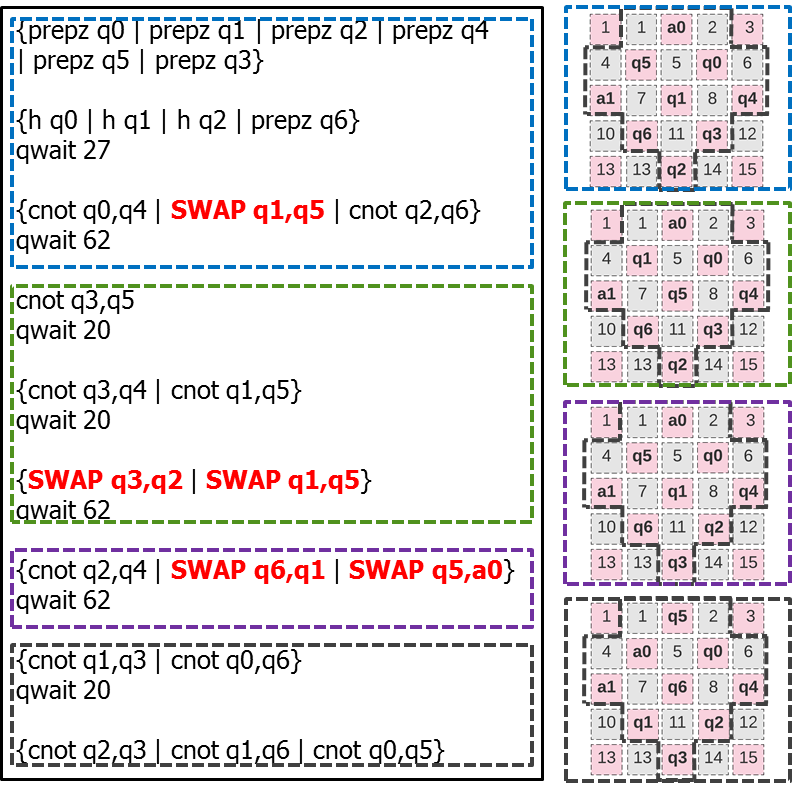}
\caption{The final circuit and the intermediate qubit layouts after mapping the Steane $[[7,1,3]]$ encoding circuit onto the checkerboard architecture.}
\label{mapcarch_7enc}
\end{figure}

The results of routing the Steane $[[7,1,3]]$ encoding circuit onto the tile-based and checkerboard architectures are shown in Figure~\ref{maptarch_7enc} and Figure~\ref{mapcarch_7enc}, respectively. 
The inputs of the routing process include 1) the pre-scheduled circuit using an ALAP approach in Figure~\ref{steane_qasm_para}; 2) the initial placement in a pre-defined architecture in Figure~\ref{iplace_7enc}.
The routing process selects the communication path and inserts SWAP operations when two qubits for a coming CNOT gate are not neighbours and then the qubit layout is changed.
The final circuits and the intermediate qubit layouts after a full mapping procedure on the tile-based architecture and the checkerboard architecture are shown in Figures~\ref{maptarch_7enc} and \ref{mapcarch_7enc} respectively. Note that the operations inside each dashed block will be executed on the qubit layout marked in the same color and the current layout will be transformed into the next one after performing the inserted SWAP operation(s).
Note that the final circuits after routing are totally different from the original circuit with an ALAP scheduling.
This is because the operations inside each routing buffer has been rescheduled using an ASAP approach.

\section{Metrics and benchmarks}
\label{sec-metric}

In order to evaluate the impact of the mapping passes as well as the proposed qubit plane architectures we define the following metrics:


\textbf{Qubit efficiency $E_{q}$:} It is calculated as $E_{q}=\frac{\# Data}{\# AllQubits}$; where \# AllQubits refers to the total number of logical qubits in a predefined qubit architecture for executing a quantum algorithm, including both logical data qubits and logical ancilla qubits, and \# Data is the number of logical data qubits. 

\textbf{Circuit latency:} It is the total execution time of a quantum algorithm in SC cycles. 
Even though reducing the circuit latency may have an overall negligible impact on the exponential performance improvement, it may be important for the algorithms with polynomial speedup. 
More importantly, shorter latency will also decrease the failure rate of the executed circuit.

\textbf{Latency overhead:} It is the percentage of latency used for moving qubits, and it is calculated as  $\frac{L_{R}-L_{S}}{L_{S}}$; where $L_{R}$ and $L_{S}$ are the circuit latency with and without considering routing qubits, respectively.

\textbf{Operation overhead:} It is the percentage of inserted movement operations and it is calculated as $\frac{\# SWAPs}{\# Gates}$; where \# Gates is the number of operations of the quantum algorithm which has not been routed (see Table \ref{qa_benchmark}) and
\# SWAPs is the total number of $\swap$ operations that are inserted for routing qubits. 
Reducing the number of operations for qubit communication helps to improve the computation fidelity. 

\textbf{Communication overhead:} It is expressed in terms of both operation overhead and latency overhead.

The benchmarks used for this mapping evaluation are shown in Table~\ref{qa_benchmark} from Qlib~\cite{lin2014qlib} and RevLib~\cite{miller2003transformation}. 
These circuits are decomposed into ones which only contain the gates from the fault-tolerantly implementable universal set $\{$Pauli, $H, \cnot, S, T\}$ on surface codes. 
We characterize these benchmarks in terms of percentage of CNOT gates $R_{cg}=\frac{\# CNOTs}{\# Gates}$, percentage of edges which have name dependency ($E^{2}_{G}$) in the QODG $R_{cd} = \frac{\left|E^{2}_{G}\right|}{\left|E_{G}\right|}$, and percentage of expensive $T, T^{\dagger}$ and $S, S^{\dagger}$ gates $R_{tsg}=\frac{(\# Ss+\#Ts+\# S^{\dagger}s+\#T^{\dagger}s)}{\# Gates}$. 
The first two benchmarks are encoding circuits of different QEC codes which are used for preparing magic states on SC \cite{fowler2012surface}. 
Table ~\ref{qa_benchmark} also shows the size ($R\times C$) of a qubit plane architecture, where $R$ and $C$ represent the number of data qubits in the $x$ axis and $y$ axis of the defined qubit plane architecture, respectively.

\begin{table}[bt!]
\centering
\caption{Quantum algorithm benchmarks}
\label{qa_benchmark}
\small
\resizebox{0.47\textwidth}{!}{
\begin{tabular}{|c|c|c|c|c|c|c|c|}
\hline
Benchmarks & \# Qubits & \# Gates & \#CNOT & Rcg\% & Rcd\% & Rtsg\%                    & Size                     \\ \hline
7-enc      & 7         & 21       & 12     & 52.38 & 42.55 & 0                         & 3$\times$3                      \\ \hline
15-enc     & 15        & 53       & 35     & 64.15 & 60.17 & 0                         & 4$\times$4                      \\ \hline
Adder0-5   & 16        & 306      & 126    & 41.18 & 26.1  & 48.0                      & 4$\times$4                      \\ \hline
Adder1-8     & 18        & 289      & 129    & 44.64 & 22.38 & 45.3                      & 5$\times$5                      \\ \hline
Adder1-16    & 34        & 577      & 257    & 44.54 & 22.16 & 44.9                      & 6$\times$6 \\ \hline
Multiply4       & 21        & 1655     & 722    & 43.63 & 18.20 & 44.4                      & 5$\times$5 \\ \hline
Shor15     & 11        & 4792     & 1788   & 37.31 & 21.03 & 48.4                      & 4$\times$3                      \\ \hline
sqrt7      & 15        & 7630     & 3089   & 40.48 & 6.41  & 43.72                     & 4$\times$4                      \\ \hline
sqrt8      & 12        & 3009     & 1314   & 43.67 & 4.63  & 43.50                     & 4$\times$3                      \\ \hline
ham7       & 7         & 320      & 149    & 35.63 & 5.67  & 41.56                     & 3$\times$3                      \\ \hline
hwb5       & 5         & 233      & 107    & 45.92 & 5.52  & 42.06                     & 3$\times$2                      \\ \hline
hwb6       & 6         & 1336     & 598    & 44.76 & 5.26  & 42.96                     & 3$\times$2                      \\ \hline
hwb7       & 7         & 6723     & 2952   & 61.60 & 4.64  & 43.62                     & 3$\times$3                      \\ \hline
rd73       & 10        & 230      & 104    & 45.22 & 4.61  & 42.61                     & 4$\times$3                      \\ \hline
rd84       & 15        & 343      & 154    & 44.90 & 4.63  & 42.86                     & 4$\times$4                      \\ \hline
\end{tabular}}
\end{table}

\section{Results}
\label{sec-result}

\begin{figure*}[tb!]
    \centering
    \includegraphics[width=\textwidth]{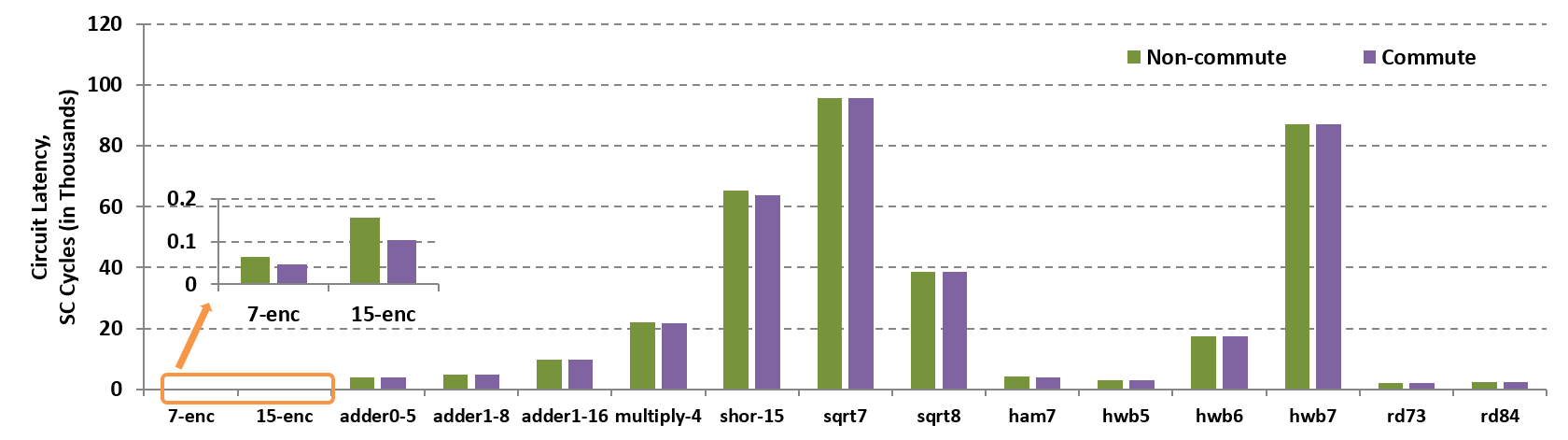}
    \caption{Comparison of the scheduling models with and without considering the commutation property ($d=3$).}
    \label{commutationd3}
\end{figure*}

\begin{figure*}[tb!]
    \centering
    \includegraphics[width=\textwidth]{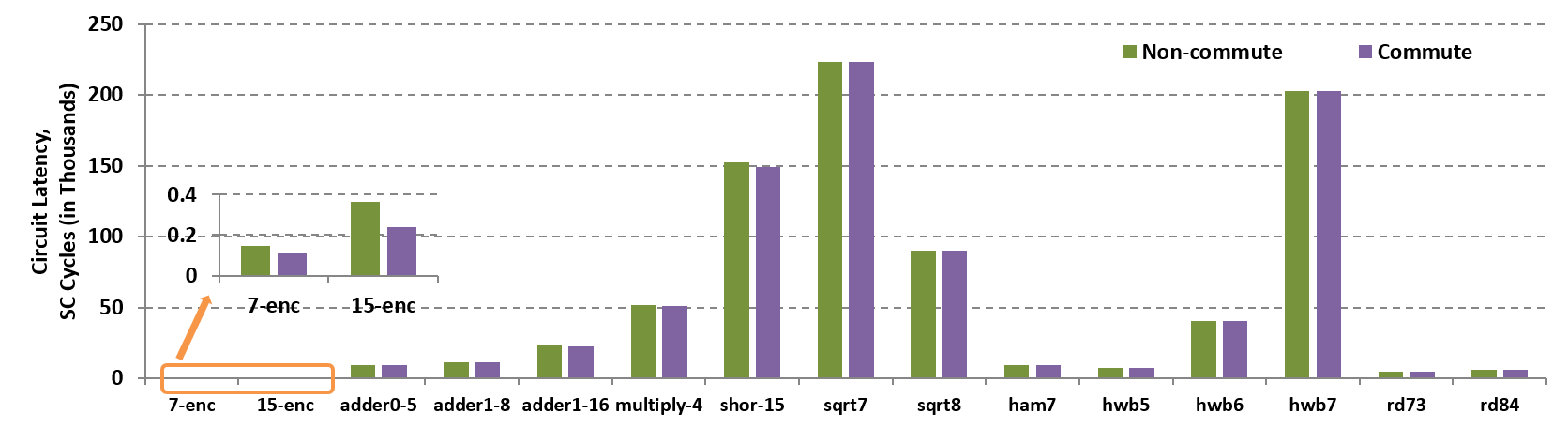}
    \caption{Comparison of the scheduling models with and without considering the commutation property ($d=7$).}
    \label{commutationd7}
\end{figure*}

We map the benchmarks shown in Table~\ref{qa_benchmark} onto the two introduced qubit architectures using the proposed mapping passes. 
As shown in Table~\ref{time_log}, the execution time of different operations is determined by the code distance $d$ which is a tunable parameter of the mapping procedure. 
In this section, only the mapping results for distance-$3$ and distance-$7$ planar SC are presented,
the results for other distances will be similar. 

We first analyze the impact of the CNOT commutation property (Section~\ref{sec-schedule}) on the latency of scheduled quantum circuits. 
We only show the results for the ALAP scheduling as they are similar to the ASAP scheduling. 
Figures~\ref{commutationd3} ( \ref{commutationd7}) compares the proposed scheduling models for distance $3$ (distance $7$) with and without taking the commutation property into account. 
For the encoding circuits, the scheduling considering commutation can significantly reduce the circuit latency, $28.1\%$ ($23.7\%$) for 7-enc and $34.4\%$ ($34.5\%$) for 15-enc, compared to the scheduling without considering commutation. 
This is because they have a high percentage of commutable $\cnot$ gates ($R_{cd}$) meanwhile the percentage of expensive gates ($R_{tsg}$) is much lower ($0$). 
In contrast, for the other benchmarks the benefit of considering commutation is negligible, up to $\sim 4\%$ ($\sim 4\%$) for adder0-5. 

\begin{figure*}[tbh!]
    \centering
    \includegraphics[width=\textwidth]{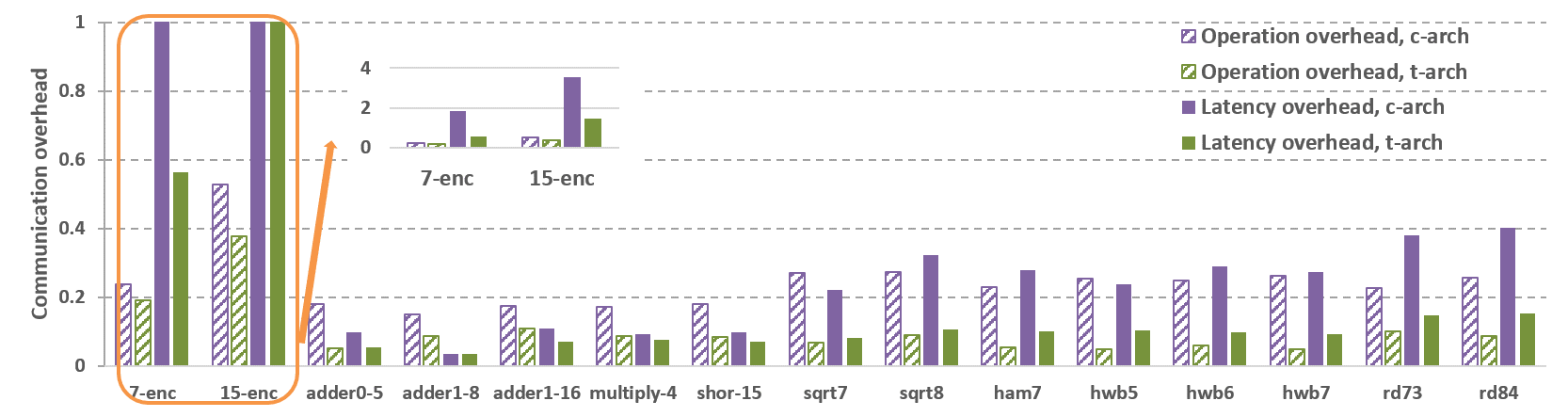}
    \caption{Comparison of mapping FT circuits onto different qubit architectures ($d=3$). The latency overhead for the first two circuits are larger than $1$ as shown in the sub-figure.}
    \label{arch_comd3}
\end{figure*}

\begin{figure*}[tbh!]
    \centering
    \includegraphics[width=\textwidth]{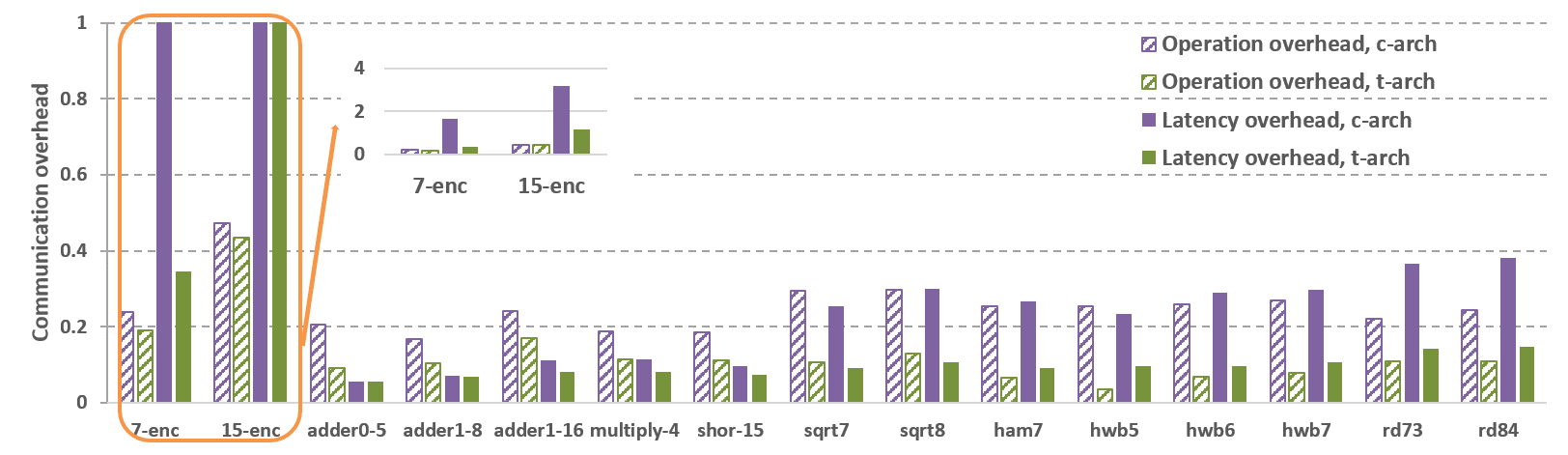}
    \caption{Comparison of mapping FT circuits onto different qubit architectures ($d=7$).}
    \label{arch_comd7}
\end{figure*}

Furthermore, we perform the full mapping procedure proposed in Section~\ref{sec-mapping}, including scheduling, placement and routing, on both the tile-based architecture (t-arch) and the checkerboard architecture (c-arch). 
The scheduling is implemented by the ALAP approach with considering commutation property.
The initial placement is achieved by either the smart approach based on Manhattan distance or the naive method which places qubits in order.
Note that the effect of initial placement is not always important \cite{lin2015paqcs}, depending on the benchmarks (see Appendix~\ref{app:placement}).
In this section, the best mapping result of the above two placement approaches for each benchmark is chosen. 

\textbf{Communication overhead:}
As mentioned previously, the mapping process results in an increase of the number of quantum operations (operation overhead) as well as in an increase in the circuit latency (latency overhead). 
We evaluate the communication overhead of mapping quantum circuits on different qubit plane architectures, namely the tile-based architecture (t-arch) and the checkerboard architecture (c-arch).
Figure~\ref{arch_comd3} and \ref{arch_comd7} show the comparison between t-arch and c-arch for distance $3$ and $7$ surface codes, respectively. 
The mapping results for both distances are similar and the t-arch achieves less communication overhead because it has a higher number of nearest neighbours. 

The operation overhead in the t-arch compared to the c-arch is reduced by $20.0\%$ (7-enc) up to $81.4\%$ (hwb5) for $d=3$ and by $8.0\%$ (15-enc) up to $86.4\%$ (hwb5) for $d=7$ . 
The latency overhead when mapping on the t-arch shows a reduction of $3.2\%$ (adder1-8), $16.3\%$ (multiply-4) and up to $69.1\%$ (7-enc) for distance 3. And $1.0\%$ (adder1-8), $25.6\%$ (shor-15) and up to $79.4\%$ (7-enc) for distance 7.  
Note that this latency reduction is not only due to the less number of movement operations but also due to the use of much faster movements (t-SWAP) although the $\cnot$ gates (t-CNOT) are slightly slower. 

\textbf{Qubit efficiency:}
As mentioned in Section~\ref{sec-arch}, $1/4$ and $\sim 1/2$ of the total number of patches are used for allocating logical data qubits in the tile-based architecture and the checkerboard architecture, respectively.
Therefore, the qubit efficiency in the t-arch is $E_{q}=1/4$ and the qubit efficiency in the c-arch is $E_{q}\approx 1/2$.

Based on the above observations, we can conclude that although the tile-based architecture is less qubit efficient than the checkerboard architecture, it can also substantially reduce the communication overhead in terms of operation overhead (up to $\sim 86\%$) and latency overhead (up to $\sim 79\%$). 
As we mentioned previously, decreasing the communication overhead helps to improve the computation fidelity. 
Therefore, one may have to compromise between qubit efficiency and communication overhead for the realization of quantum algorithms.

\section{Conclusion}
\label{sec-conc}

We have proposed two SC qubit plane architectures to efficiently support the execution of lattice surgery-based quantum circuits. We developed a full procedure for mapping small-scale quantum algorithms onto these two SC architectures. 
The experimental results show the following observations. 
First, the proposed scheduling considering the commutation property provides faster circuit execution than the scheduling without considering commutation. 
Secondly, the mapping procedure causes communication overhead in terms of both operation overhead and latency overhead.
Moreover, the communication overhead highly depends on how qubits are organized and moved, that is, the qubit plane architectures. 
The tile-based architecture considerably decreases the number of movements and also supports faster execution compared to the checkerboard though it is less qubit-efficient. 

As future work, we will focus on heuristic scheduling and placement algorithms as well as different routing techniques for large-scale quantum benchmarks. 
Furthermore, we will consider the dynamics of quantum computation such as magic state distillation for $S$ or $T$ gates and qubit routing for performing `neighbouring' CNOT gates.
Then we will investigate their implications on quantum circuit mapping.
In addition, we will investigate different qubit architectures, for instance, an architecture with specialized communication channels for moving qubits and pre-defined regions for preparing magic states.

\bibliographystyle{ieeetr}
\bibliography{References} 

\begin{thebibliography}{10}

\bibitem{shor1994algorithms}
P.~W. Shor, ``Algorithms for quantum computation: Discrete logarithms and
  factoring,'' in {\em SFCS}, 1994.

\bibitem{barends2014superconducting}
R.~Barends {\em et~al.}, ``Superconducting quantum circuits at the surface code
  threshold for fault tolerance,'' {\em Nature}, vol.~508, no.~7497,
  pp.~500--503, 2014.

\bibitem{versluis2016scalable}
R.~Versluis {\em et~al.}, ``Scalable quantum circuit and control for a
  superconducting surface code,'' {\em arXiv:1612.08208}, 2016.

\bibitem{hill2015surface}
C.~D. Hill {\em et~al.}, ``A surface code quantum computer in silicon,'' {\em
  Science advances}, vol.~1, no.~9, p.~e1500707, 2015.

\bibitem{li2017crossbar}
R.~Li {\em et~al.}, ``A crossbar network for silicon quantum dot qubits,'' {\em
  arXiv:1711.03807}, 2017.

\bibitem{IBM17}
IBM, ``Quantum experience.''

\bibitem{boixo2016characterizing}
S.~Boixo {\em et~al.}, ``Characterizing quantum supremacy in near-term
  devices,'' {\em arXiv:1608.00263}, 2016.

\bibitem{sete2016functional}
E.~A. Sete {\em et~al.}, ``A functional architecture for scalable quantum
  computing,'' in {\em ICRC}, pp.~1--6, IEEE, 2016.

\bibitem{bishop2017quantum}
L.~S. Bishop {\em et~al.}, ``Quantum volume,'' tech. rep., 2017.

\bibitem{linke2017experimental}
N.~M. Linke {\em et~al.}, ``Experimental comparison of two quantum computing
  architectures,'' {\em Proceedings of the National Academy of Sciences},
  p.~201618020, 2017.

\bibitem{metodi2006scheduling}
T.~S. Metodi {\em et~al.}, ``Scheduling physical operations in a quantum
  information processor,'' in {\em SPIE}, 2006.

\bibitem{whitney2007automated}
M.~Whitney {\em et~al.}, ``Automated generation of layout and control for
  quantum circuits,'' in {\em CF}, 2007.

\bibitem{dousti12min}
M.~J. Dousti and M.~Pedram, ``Minimizing the latency of quantum circuits during
  mapping to the ion-trap circuit fabric,'' in {\em DATE}, 2012.

\bibitem{yazdani2013quantum}
M.~Yazdani {\em et~al.}, ``A quantum physical design flow using ilp and graph
  drawing,'' {\em Quantum information processing}, vol.~12, no.~10,
  pp.~3239--3264, 2013.

\bibitem{bahreini2015minlp}
T.~Bahreini and N.~Mohammadzadeh, ``An minlp model for scheduling and placement
  of quantum circuits with a heuristic solution approach,'' {\em JETC},
  vol.~12, no.~3, p.~29, 2015.

\bibitem{lye2015determining}
A.~Lye {\em et~al.}, ``Determining the minimal number of swap gates for
  multi-dimensional nearest neighbor quantum circuits,'' in {\em ASP-DAC},
  pp.~178--183, IEEE, 2015.

\bibitem{wille2016look}
R.~Wille {\em et~al.}, ``Look-ahead schemes for nearest neighbor optimization
  of 1d and 2d quantum circuits,'' in {\em ASP-DAC}, pp.~292--297, IEEE, 2016.

\bibitem{farghadan2017quantum}
A.~Farghadan and N.~Mohammadzadeh, ``Quantum circuit physical design flow for
  2d nearest-neighbor architectures,'' {\em International Journal of Circuit
  Theory and Applications}, vol.~45, no.~7, pp.~989--1000, 2017.

\bibitem{IBMQISKIT}
IBM, ``Qiskit, quantum information software kit.''

\bibitem{Zulehner2017efficient}
A.~Zulehner {\em et~al.}, ``An efficient methodology for mapping quantum
  circuits to the ibm qx architectures,'' {\em arXiv:1712.04722}, 2017.

\bibitem{siraichi2018qubit}
M.~Siraichi {\em et~al.}, ``Qubit allocation,'' in {\em ACM-CGO}, pp.~1--12,
  2018.

\bibitem{venturelli2018compiling}
D.~Venturelli, M.~Do, E.~Rieffel, and J.~Frank, ``Compiling quantum circuits to
  realistic hardware architectures using temporal planners,'' {\em Quantum
  Science and Technology}, vol.~3, no.~2, p.~025004, 2018.

\bibitem{riste2015detecting}
D.~Rist{\`e} {\em et~al.}, ``Detecting bit-flip errors in a logical qubit using
  stabilizer measurements,'' {\em Nature communications}, vol.~6, no.~6983,
  2015.

\bibitem{kelly2015state}
J.~Kelly {\em et~al.}, ``State preservation by repetitive error detection in a
  superconducting quantum circuit,'' {\em Nature}, vol.~519, no.~7541,
  pp.~66--69, 2015.

\bibitem{nielsen2010quantum}
M.~A. Nielsen and I.~L. Chuang, {\em Quantum computation and quantum
  information}.
\newblock Cambridge university press, 2010.

\bibitem{balensiefer2005quale}
S.~Balensiefer {\em et~al.}, ``Quale: quantum architecture layout evaluator,''
  in {\em SPIE}, 2005.

\bibitem{dousti2013leqa}
M.~J. Dousti and M.~Pedram, ``Leqa: latency estimation for a quantum algorithm
  mapped to a quantum circuit fabric,'' in {\em DAC}, 2013.

\bibitem{dousti2014squash}
M.~J. Dousti {\em et~al.}, ``Squash: a scalable quantum mapper considering
  ancilla sharing,'' in {\em GLSVLSI}, 2014.

\bibitem{ahsan2015architecture}
M.~Ahsan, {\em Architecture Framework for Trapped-Ion Quantum Computer based on
  Performance Simulation Tool}.
\newblock PhD thesis, Duke University, 2015.

\bibitem{heckey2015compiler}
J.~Heckey {\em et~al.}, ``Compiler management of communication and parallelism
  for quantum computation,'' in {\em ASPLOS}, 2015.

\bibitem{lin2015paqcs}
C.-C. Lin {\em et~al.}, ``Paqcs: Physical design-aware fault-tolerant quantum
  circuit synthesis,'' {\em IEEE Transactions on VLSI Systems}, vol.~23, no.~7,
  pp.~1221--1234, 2015.

\bibitem{bravyi1998quantum}
S.~B. Bravyi and A.~Y. Kitaev, ``Quantum codes on a lattice with boundary,''
  {\em arXiv:9811052}, 1998.

\bibitem{paler2016synthesis}
A.~Paler {\em et~al.}, ``Synthesis of arbitrary quantum circuits to topological
  assembly,'' {\em Scientific reports}, vol.~6, p.~30600, 2016.

\bibitem{paler2017fault}
A.~Paler {\em et~al.}, ``Fault-tolerant, high-level quantum circuits: form,
  compilation and description,'' {\em Quantum Science and Technology}, vol.~2,
  no.~2, p.~025003, 2017.

\bibitem{paler2017online}
A.~Paler {\em et~al.}, ``Online scheduled execution of quantum circuits
  protected by surface codes,'' {\em arXiv:1711.01385}, 2017.

\bibitem{javadi2017optimized}
Javadi-Abhari {\em et~al.}, ``Optimized surface code communication in
  superconducting quantum computers,'' in {\em MICRO}, pp.~692--705, ACM, 2017.

\bibitem{horsman2012surface}
C.~Horsman {\em et~al.}, ``Surface code quantum computing by lattice surgery,''
  {\em New Journal of Physics}, vol.~14, no.~12, p.~123011, 2012.

\bibitem{landahl2014quantum}
A.~J. Landahl and C.~Ryan-Anderson, ``Quantum computing by color-code lattice
  surgery,'' {\em arXiv preprint arXiv:1407.5103}, 2014.

\bibitem{herr2017optimization}
D.~Herr {\em et~al.}, ``Optimization of lattice surgery is np-hard,'' {\em npj
  Quantum Information}, vol.~3, no.~1, p.~35, 2017.

\bibitem{dennis2002topological}
E.~Dennis {\em et~al.}, ``Topological quantum memory,'' {\em Journal of
  Mathematical Physics}, vol.~43, no.~9, pp.~4452--4505, 2002.

\bibitem{raussendorf2006fault}
R.~Raussendorf {\em et~al.}, ``A fault-tolerant one-way quantum computer,''
  {\em Annals of physics}, vol.~321, no.~9, pp.~2242--2270, 2006.

\bibitem{fowler2012surface}
A.~G. Fowler {\em et~al.}, ``Surface codes: Towards practical large-scale
  quantum computation,'' {\em Physical Review A}, vol.~86, no.~3, p.~032324,
  2012.

\bibitem{raussendorf2007fault}
R.~Raussendorf and J.~Harrington, ``Fault-tolerant quantum computation with
  high threshold in two dimensions,'' {\em Physical review letters}, vol.~98,
  no.~19, p.~190504, 2007.

\bibitem{raussendorf2007topological}
R.~Raussendorf, J.~Harrington, and K.~Goyal, ``Topological fault-tolerance in
  cluster state quantum computation,'' {\em New Journal of Physics}, vol.~9,
  no.~6, p.~199, 2007.

\bibitem{bravyi2005universal}
S.~Bravyi and A.~Kitaev, ``Universal quantum computation with ideal clifford
  gates and noisy ancillas,'' {\em Physical Review A}, vol.~71, no.~2,
  p.~022316, 2005.

\bibitem{bravyi2012magic}
S.~Bravyi and J.~Haah, ``Magic-state distillation with low overhead,'' {\em
  Physical Review A}, vol.~86, no.~5, p.~052329, 2012.

\bibitem{meier2012magic}
A.~M. Meier, B.~Eastin, and E.~Knill, ``Magic-state distillation with the
  four-qubit code,'' {\em arXiv:1204.4221}, 2012.

\bibitem{jones2013multilevel}
C.~Jones, ``Multilevel distillation of magic states for quantum computing,''
  {\em Physical Review A}, vol.~87, no.~4, p.~042305, 2013.

\bibitem{campbell2017unifying}
E.~T. Campbell and M.~Howard, ``Unifying gate synthesis and magic state
  distillation,'' {\em Physical Review L}, vol.~118, no.~6, p.~060501, 2017.

\bibitem{gottesman1998fault}
D.~Gottesman, ``Fault-tolerant quantum computation with higher-dimensional
  systems,'' {\em arXiv:9802007}, 1998.

\bibitem{fu2016heterogeneous}
X.~Fu {\em et~al.}, ``A heterogeneous quantum computer architecture,'' in {\em
  CF}, 2016.

\bibitem{hennessy2011computer}
J.~L. Hennessy and D.~A. Patterson, {\em Computer architecture: a quantitative
  approach}.
\newblock Elsevier, 2011.

\bibitem{shafaei2014qubit}
A.~Shafaei {\em et~al.}, ``Qubit placement to minimize communication overhead
  in 2d quantum architectures,'' in {\em ASP-DAC}, 2014.

\bibitem{richards2002spacecraft}
A.~Richards {\em et~al.}, ``Spacecraft trajectory planning with avoidance
  constraints using mixed-integer linear programming,'' {\em JGCD}, vol.~25,
  no.~4, pp.~755--764, 2002.

\bibitem{kaufman1978algorithm}
L.~Kaufman and F.~Broeckx, ``An algorithm for the quadratic assignment problem
  using bender's decomposition,'' {\em EJOR}, vol.~2, no.~3, pp.~207--211,
  1978.

\bibitem{lin2014qlib}
C.-C. Lin {\em et~al.}, ``Qlib: Quantum module library,'' {\em ACM-JETC},
  vol.~11, no.~1, p.~7, 2014.

\bibitem{miller2003transformation}
D.~M. Miller {\em et~al.}, ``A transformation based algorithm for reversible
  logic synthesis,'' in {\em DAC}, pp.~318--323, IEEE, 2003.

\bibitem{gottesman1998heisenberg}
D.~Gottesman, ``The heisenberg representation of quantum computers,'' {\em
  arXiv:9807006}, 1998.

\end{thebibliography}

\clearpage
\begin{appendices}
\section{Lattice surgery-based CNOT}
\label{app:cnot}

A CNOT is a gate applying on two qubits, the target qubit undergoes an $X$ gate only if the control qubit is in $\ket{1}$. One way to validate a CNOT implementation is to check the transformation of logical $X$ and $Z$ operators using the Heisenberg representation~\cite{gottesman1998heisenberg} as follows:
\begin{equation}
\label{equ:xi}
    CNOT^{\dagger}(X\otimes I) CNOT = X \otimes X
\end{equation}
\begin{equation}
\label{equ:ix}
    CNOT^{\dagger}(I\otimes X) CNOT = I \otimes X
\end{equation}
\begin{equation}
\label{equ:zi}
    CNOT^{\dagger}(Z\otimes I) CNOT = Z \otimes I
\end{equation}
\begin{equation}
\label{equ:iz}
    CNOT^{\dagger}(I\otimes Z) CNOT = Z \otimes Z
\end{equation}
For instance, the CNOT gate transforms an $X$ in the control qubit into the target qubit in Equation (\ref{equ:xi}).
We can verify the measurement-based procedure ~\cite{gottesman1998fault}, which is described by the circuits in Figure~\ref{cnot_circuit1} and \ref{cnot_circuit2}, by examining these transformations ((\ref{equ:xi})-(\ref{equ:iz})) as shown in Equations (\ref{equ:cnot1}) and (\ref{equ:cnot2}) respectively. 
These equations illustrate how different measurements transform stabilizers and logical operators.
`C', `T', and `A' represent the control, target, and ancillary qubit, respectively. 
`S' and `L' represent the stabilizers and logical operators, respectively.
For example, after performing measurements $M_{IXX}$ in (\ref{equ:cnot1}), the stabilizer $IZI$ is transformed into $(-1)^{M_{IXX}}IXX$ and the logical operator $IIZ$ is transformed into $IZZ$.
Equations (\ref{equ:cnot1}) and (\ref{equ:cnot2}) show that the measurement-based procedure does satisfy the transform relations in Equations (\ref{equ:xi})-(\ref{equ:iz}) and it is thus equivalent to a CNOT.

\begin{equation}
\label{equ:cnot1}
\resizebox{0.5\textwidth}{!}{$\begin{array}{c|ccrcrcr}
\multicolumn{1}{l|}{} & \multicolumn{4}{l}{CAT}                             \\ \hline
S                      & IZI & & (-1)^{M_{IXX}}IXX & &(-1)^{M_{ZZI}}ZZI & &(-1)^{M_{IXI}}IXI \\ 
L                      & XII & &XII            && (-1)^{M_{IXX}} XXX           && (-1)^{M_{IXX}+M_{IXI}} XIX           \\ 
                       & ZII &\overset{M_{IXX}}{\rightarrow}& ZII            &\overset{M_{ZZI}}{\rightarrow}& ZII           &\overset{M_{IXI}}{\rightarrow}& ZII           \\ 
                       & IIX && IIX            && IIX           && IIX           \\ 
                       & IIZ && IZZ            && IZZ           && (-1)^{M_{ZZI}} ZIZ           \\ 
\end{array}$}
\end{equation}
\begin{equation}\label{equ:cnot2}
\resizebox{0.5\textwidth}{!}{$\begin{array}{c|ccrcrcr}
\multicolumn{1}{l|}{} & \multicolumn{4}{l}{CAT}                             \\ \hline
S                      & IXI & & (-1)^{M_{ZZI}}ZZI & &(-1)^{M_{IXX}}IXX & &(-1)^{M_{IZI}}IZI \\ 
L                      & XII & &XXI            && XXI           && (-1)^{M_{IXX}} XIX           \\ 
                       & ZII &\overset{M_{ZZI}}{\rightarrow}& ZII            &\overset{M_{IXX}}{\rightarrow}& ZII           &\overset{M_{IZI}}{\rightarrow}& ZII           \\ 
                       & IIX && IIX            && IIX           && IIX           \\ 
                       & IIZ && IIZ            && (-1)^{M_{ZZI}}ZZZ           && (-1)^{M_{ZZI}+M_{IZI}} ZIZ           \\ 
\end{array}$}
\end{equation}


The joint measurement $M_{XX}$ ($M_{ZZ}$) is realized by merge and split operations using lattice surgery \cite{horsman2012surface, landahl2014quantum}. The basic operations of lattice surgery are to stop measuring some existing stabilizers and start to measure some new stabilizers. 
For example, the merge operation for $M_{ZZ}$ on the qubits `A' and `C' in Figure~\ref{layout_cnot} is performed by ceasing to measure $X_{7}X_{8}$ and $X_{10}X_{11}$, starting to measure $Z_{6}Z_{7}Z_{9}Z_{10}$, $Z_{8}Z_{11}$ and $X_{7}X_{8}X_{10}X_{11}$, that is, performing $d$ rounds of ESM on the merged lattice in Figure~\ref{merge}. This means the two lattices `$A$' and `$C$' are integrated into one single lattice. Similarly, the split operation is implemented by ceasing to measure $Z_{6}Z_{7}Z_{9}Z_{10}$, $Z_{8}Z_{11}$ and $X_{7}X_{8}X_{10}X_{11}$, starting to measure stabilizer $X_{7}X_{8}$ and $X_{10}X_{11}$ , that is, performing $d$ rounds of ESM individually on each lattice `$A$' and `$C$' in Figure~\ref{split}. The splitting procedure divides the merged lattice back into two lattices. Afterwards, one needs to read out the outcome of each joint measurement for further logical Pauli corrections.
The measurement result of $M_{ZZ}$ is interpreted into $0$ ($1$) if the number of `$-$' syndromes from the new stabilizers $Z_{6}Z_{7}Z_{9}Z_{10}$ and $Z_{8}Z_{11}$ during the merge is even (odd). 


\section{Lattice surgery-based movement}
\label{app:movement}

\begin{figure}[tbh!]
\centerline{\subfigure[]{\includegraphics[width=1in]{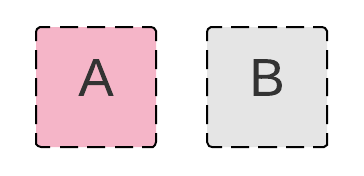}
\label{rowmove1}}
\subfigure[]{\includegraphics[width=1in]{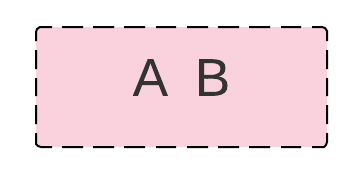}
\label{rowmove2}}
\subfigure[]{\includegraphics[width=1in]{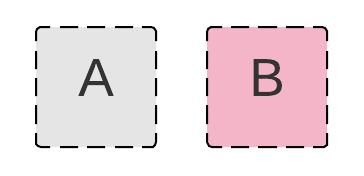}
\label{rowmove3}}}
\caption{(a) Patch A is a logical qubit in state $\ket{\psi}$ and patch B is an ancilla in state $\ket{0}$. First perform the joint measurement $M_{XX}$ realized by a merge (b) and a split (c), then perform the measurement $M_{Z}$ on patch A, the state $\ket{\psi}$ is moved to patch B.}
\label{rowmove}
\end{figure}

\begin{figure}[tbh!]
\centerline{\subfigure[]{\includegraphics[width=0.48in]{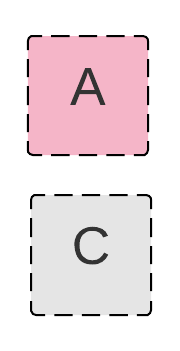}
\label{columnmove1}}
\subfigure[]{\includegraphics[width=0.48in]{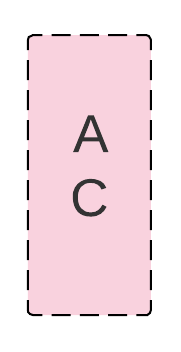}
\label{columnmove2}}
\subfigure[]{\includegraphics[width=0.48in]{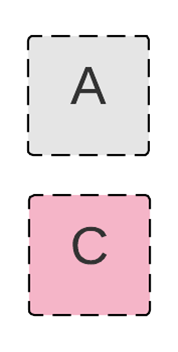}
\label{columnmove3}}}
\caption{(a) Patch A is a logical qubit in state $\ket{\psi}$ and patch C is an ancilla in state $\ket{+}$. First perform the joint measurement $M_{ZZ}$ realized by a merge (b) and a split (c), then perform the measurement $M_{X}$ on patch A, the state $\ket{\psi}$ is moved to patch C.}
\label{columnmove}
\end{figure}

\begin{figure}[tbh!]
\centerline{\subfigure[]{\includegraphics[width=0.8in]{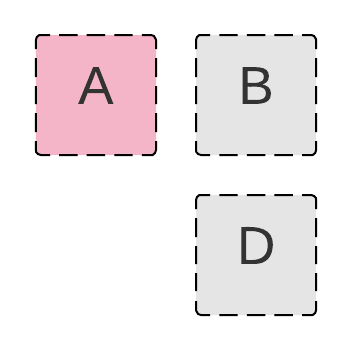}
\label{cornermove1}}
\subfigure[]{\includegraphics[width=0.84in]{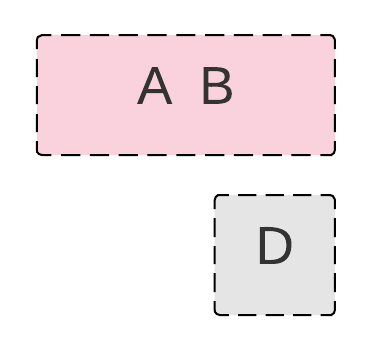}
\label{cornermove2}}
\subfigure[]{\includegraphics[width=0.86in]{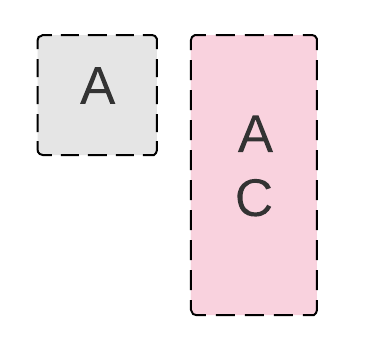}
\label{cornermove3}}
\subfigure[]{\includegraphics[width=0.8in]{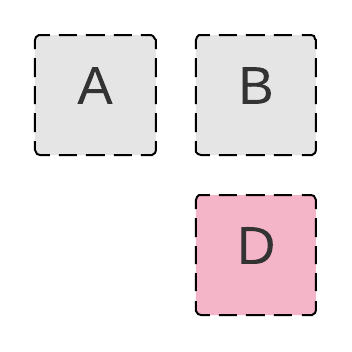}
\label{cornermove4}}}
\caption{(a) Patch A is a logical qubit in state $\ket{\psi}$, patch B and D are ancillas in states $\ket{0}$ and $\ket{+}$, respectively. The state $\ket{\psi}$ is moved to patch D in $3d$ SC cycles as follows: First perform the joint measurement $M_{ZZ}$ between A and B (b);  then perform the joint measurement $M_{XX}$ between B and D; finally perform the measurement $M_{X}$ on patch A.}
\label{cornermove}
\end{figure}

The lattice surgery-based joint measurements can be used to `move' logical qubits to other locations.
As mentioned previously, the adjacent boundaries should be in both $X$- or $Z$-type when performing such a joint measurement. Assuming that the qubit patches in the same row (column) of the tile-based architecture in Figure~\ref{tile} have $Z-$($X$-)type adjacent boundaries, we introduce two basic movements: horizontal movement (Figure~\ref{rowmove}) and vertical movement (Figure~\ref{columnmove}). A logical state in $A$ can be moved to its horizontally (vertically) adjacent position $B$ ($C$) by first performing a joint measurement $M_{XX}$ ($M_{ZZ}$) between $A$ and $B$ ($C$) followed by a $Z$ ($X$) measurement on $A$. This horizontal (vertical) movements mimics the procedure in Equation~(\ref{equ:hmove}) (Equation (\ref{equ:vmove})), that is, the logical operators in patch $A$ are transformed into patch $B$ ($C$). It means that the logical state in $A$ is moved to patch $B$ ($C$).
One can progressively move one logical state from one patch to the other by applying these horizontal movements and vertical movements as shown in Figure~\ref{cornermove}.

\begin{equation}
\label{equ:hmove}
\begin{array}{c|cccll}
  & AB &    &        &    &        \\ \hline
S & IZ &    & (-1)^{M_{XX}}XX &    & (-1)^{M_{ZI}}ZI \\
L & XI & M_{XX} & XI     & M_{ZI} & (-1)^{M_{XX}}IX \\
  & ZI &    & ZZ     &    & (-1)^{M_{ZI}}IZ
\end{array}
\end{equation}

\begin{equation}
\label{equ:vmove}
\begin{array}{c|cccll}
  & AC &    &        &    &        \\ \hline
    S & IX &    & (-1)^{M_{ZZ}}ZZ &    & (-1)^{M_{XI}}XI \\
L & XI & M_{ZZ} & XX     & M_{XI} & (-1)^{M_{XI}}IX \\
  & ZI &    & ZI    &    & (-1)^{M_{ZZ}}IZ
\end{array}
\end{equation}

\section{FT library}
\label{app:ftlib}

After the logical-level mapping, the physical-level mapping becomes trivial for several reasons. First, there is no need to place and route physical qubits since surface codes intrinsically satisfy the 2D NN constraint. Secondly, as discussed in Section~\ref{sec_ft}, each of the universal set of logical operations (preparation, measurement, Pauli, H, $\cnot$, S and T gates) on planar SC is implemented by a certain series of SC cycles. 

\begin{figure}[htb!]
    \centering
    \includegraphics[width=2.5in]{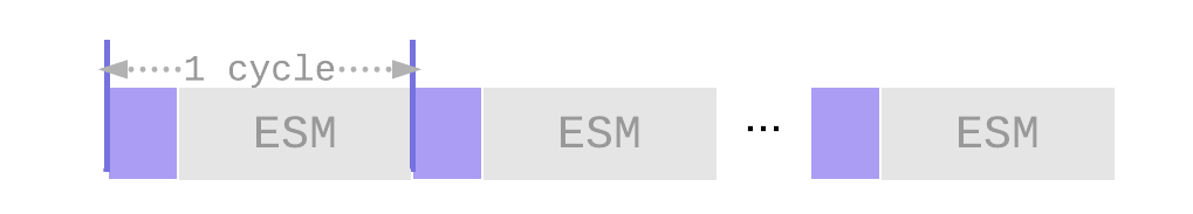}\vspace{-3mm}
    \caption{The decomposition of logical operations into SC cycles.
    }
    \label{decompose_gate}
\end{figure}

\begin{figure}[tb!]
\centering
\includegraphics[width=1.3in]{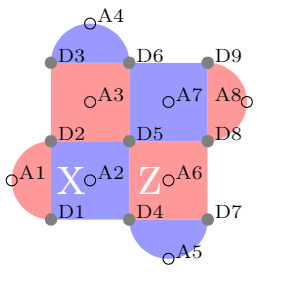}
\caption{The qubit layout for the distance-$3$ planar surface code.}
\label{ninjastar}
\end{figure}

As shown in Figure~\ref{decompose_gate}, each cycle is composed of two time slots, one purple slot for performing physical single-qubit gates and one gray slot for performing one round of ESM. Depending on the logical operation, a single-qubit gate such as Identity, Pauli gates or H gate needs to be performed during each purple slot.
For instance, a logical X gate on the distance-$3$ planar surface code (Figure~\ref{ninjastar}) can be realized by one SC cycle, that is, first performing bit-wise physical X gates on qubits $D1, D2, D3$ (purple slot) and then performing $1$ round of ESM (gray slot).
Therefore, a library can be built to translate each logical operation into pre-scheduled physical quantum operations.
Since the operations in a purple slot are bit-wise and performed in parallel, one only need to pre-schedule the operations of error syndrome measurement.

The ESM circuits for $X$ and $Z$ stabilizers are shown in Figure~\ref{esm}.
One full round of ESM on the distance-$3$ planar surface code (Figure~\ref{ninjastar}) is scheduled and performed as follows (in QASM):\\
\{ prepz A2 | prepz A7 | prepz A5\}\\
\{ h A2 | h A7 | h A5 | prepz A1 | prepz A3 | prepz A6\}\\
\{ cnot A2, D5 | cnot A7, D9 | cnot A5, D7 | cnot D2, A1 | cnot D6, A3 | cnot D8, A6 | prepz A8 | prepz A4\}\\
\{ cnot A2, D2 | cnot A7, D6 | cnot A5, D4 | cnot D9, A8 | cnot D3, A3 | cnot D5, A6 | h A4\}\\
\{ cnot A2, D4 | cnot A7, D8 | cnot A4, D6 | cnot D1, A1 | cnot D5, A3 | cnot D7, A6 | h A5\}\\
\{ cnot A2, D1 | cnot A7, D5 | cnot A4, D3 | cnot D8, A8 | cnot D2, A3 | cnot D4, A6 | measure A1 | measure A5\}\\
\{ h A2 | h A4 | h A7 | measure A3 | measure A6 | measure A8\}\\
\{ measure A2 | measure A4 | measure A7\}\\

However, a more realistic scheduling needs to consider the underlying hardware constraints such as the allowed primitive operations, their execution time, frequency multiplexing, etc. A scalable scheme for executing the ESM of surface code on superconducting qubits with NN coupling can be found in \cite{versluis2016scalable}.

\section{Initial placements}
\label{app:placement}

\begin{figure*}[tbh!]
    \centering
    \includegraphics[width=\textwidth]{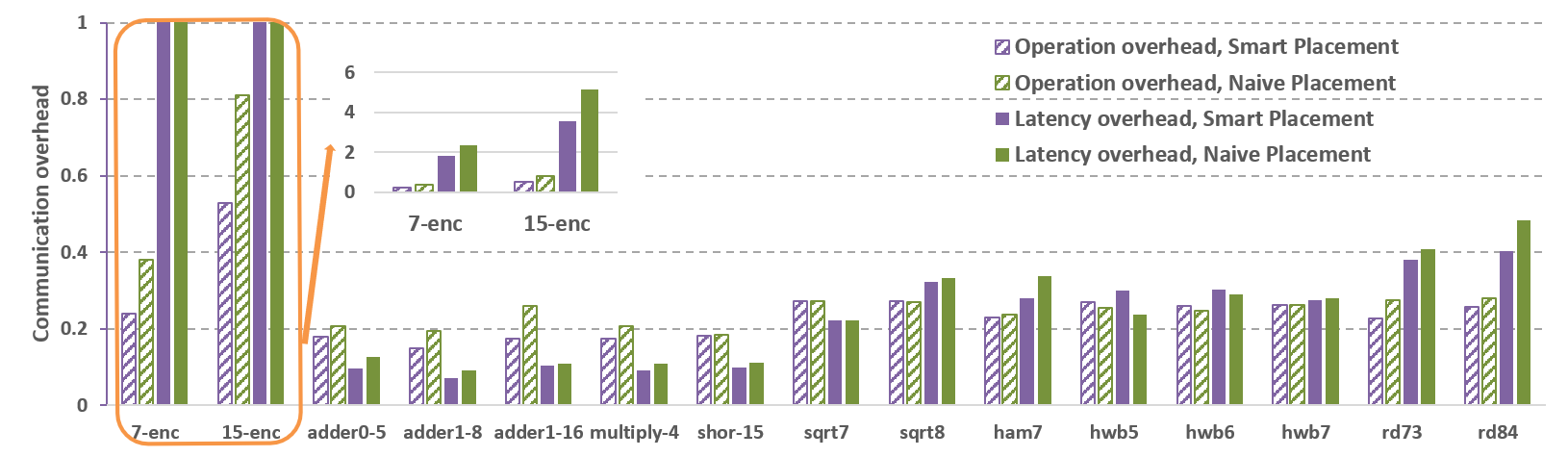}
    \caption{Comparison of mapping FT circuits with different initial placements on the checkerboard architecture ($d=3$).}
    \label{place_comcd3}
\end{figure*}
\begin{figure*}[tbh!]
    \centering
    \includegraphics[width=\textwidth]{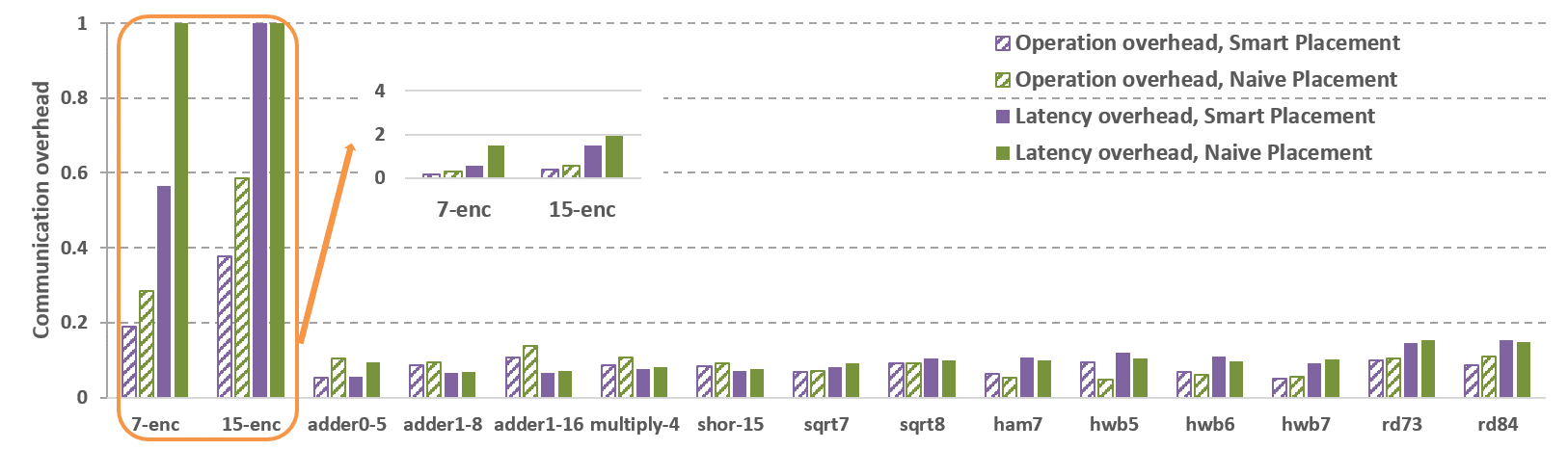}
    \caption{Comparison of mapping FT circuits with different initial placements on the tile-based architecture ($d=3$).}
    \label{place_comtd3}
\end{figure*}

\begin{figure*}[tbh!]
    \centering
    \includegraphics[width=\textwidth]{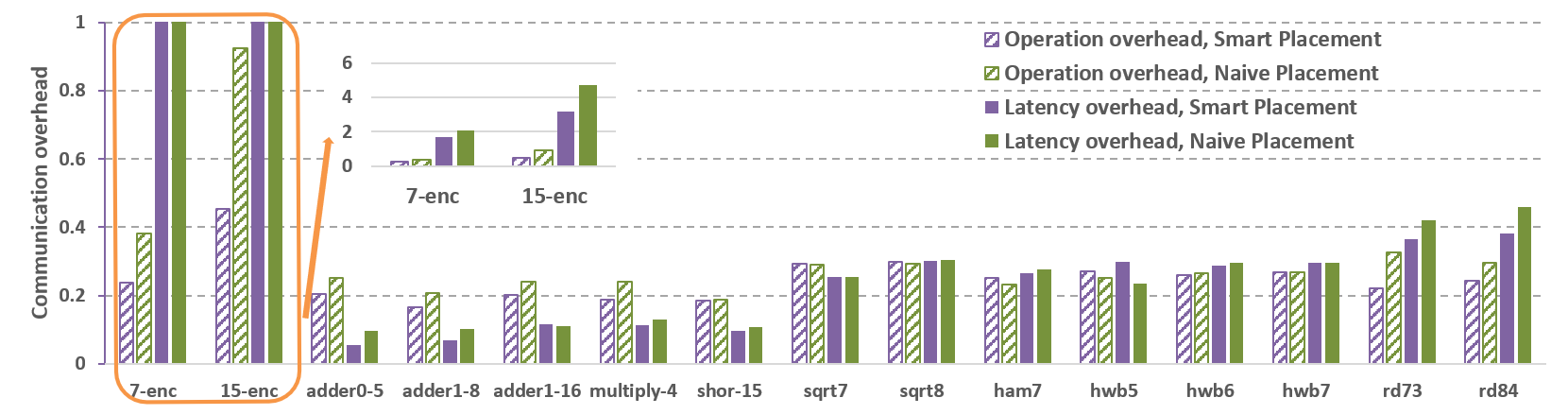}
    \caption{Comparison of mapping FT circuits with different initial placements on the checkerboard architecture ($d=7$).}
    \label{place_comcd7}
\end{figure*}

\begin{figure*}[tbh!]
    \centering
    \includegraphics[width=\textwidth]{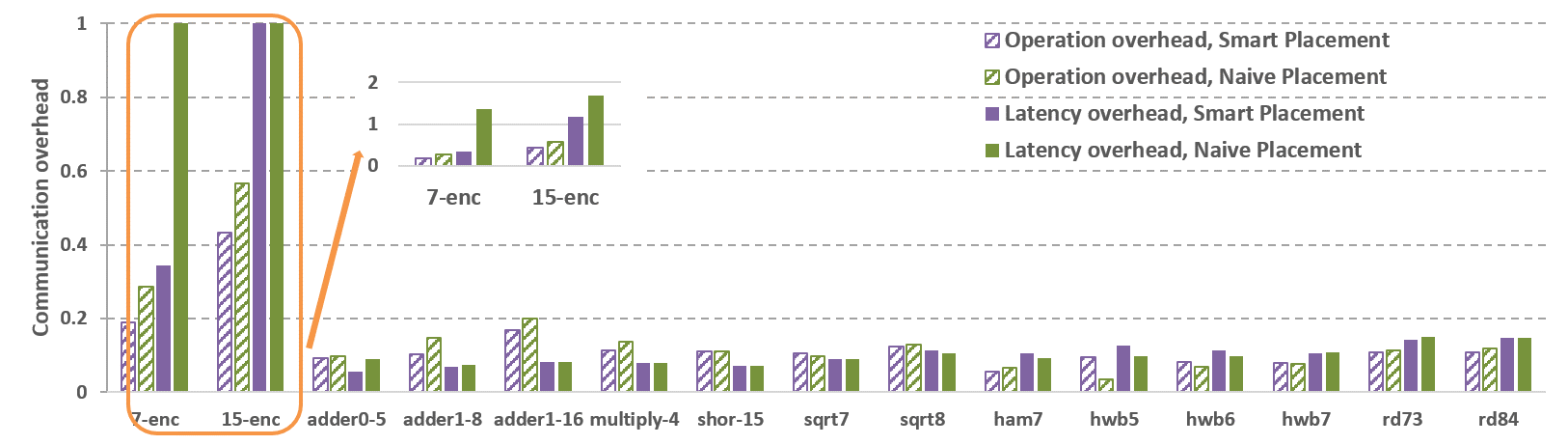}
    \caption{Comparison of mapping FT circuits with different initial placements on the tile-based architecture ($d=7$).}
    \label{place_comtd7}
\end{figure*}

In this section, we examine how initial placement affects the mapping results.

Figures~\ref{place_comcd3} and \ref{place_comtd3} show the comparison of the proposed smart placement based on Manhattan distance with a naive placement which locates qubits in order, where logical qubits are encoded by the distance-$3$ surface code. 
For some benchmarks, the use of the smart initial placement effectively decreases the operation overhead on both the c-arch and the t-arch, from $\sim 8.3\%$ up to $37.5\%$ (rd84, adder0-5, multiply-4, rd73, adder1-8, adder1-16, 15-enc) and from $\sim 7.4\%$ up to $50\%$ (adder1-8, hwb7, shor-15, multiply-4, rd84, adder1-16, 7-enc, 15-enc), respectively. 
Furthermore, the smart placement approach reduces the latency overhead of the c-arch and t-arch by $7.0\%$ to $31.1\%$ (rd73, shor-15, rd84, multiply-4, ham7, adder1-8, 7-enc, adder0-5, 15-enc) and by $7.4\%$ to $62.3\%$ (shor-15, adder1-16, hwb7, sqrt7, 15-enc, adder0-5, 7-enc), respectively. 
However, for other benchmarks, the smart placements provide marginal reductions or even increases in communication overhead on both qubit architectures. 
This is because the position of the qubits will change after each SWAP operation, and the possible benefit of the smart initial placement will progressively disappear as the circuit execution advances. 

Figures~\ref{place_comcd7} and \ref{place_comtd7} show similar results for distance-$7$ surface code. For some benchmarks, the use of smart initial placements can effectively decrease the communication overhead compared to naive placements. The smart initial placement decreases the operation overhead on the c-arch and the t-arch, from $\sim 15.8\%$ up to $51.0\%$ (adder1-16, rd84, adder0-5, adder1-8, 7-enc, 15-enc) and from $\sim 6.7\%$ up to $33.3\%$ (adder0-5, rd84, ham7, adder1-16, multiply-4, 15-enc, adder1-8, 7-enc), respectively. Moreover, the smart placement approach reduces the latency overhead of the c-arch and t-arch by $10.2\%$ to $31.1\%$ (shor-15, multiply-4, rd73, rd84, 7-enc, adder1-8, 15-enc, adder0-5) and by $5.7\%$ to $74.7\%$ (rd73, adder1-8, 15-enc, adder0-5, 7-enc), respectively. However, for other benchmarks, the benefits from smart initial placements disappear on both qubit architectures.

\end{appendices}

\end{document}